\documentclass[a4paper,11pt]{article}
\pdfoutput=1 

\usepackage{jheppub}

\usepackage[mmddyyyy,hhmmss]{datetime}

\usepackage{amsmath,graphicx,amssymb,subfigure,bbm,comment,mathtools}
\usepackage[all]{xy}

\allowdisplaybreaks

\title{Holographic three-point correlators in the Schrodinger/dipole CFT correspondence }

\author[a, b]{George Georgiou}
\author[a]{and Dimitrios Zoakos}

\affiliation[a]{Department of Physics, National and Kapodistrian University of Athens, 15784 Athens, Greece}
\affiliation[b]{Institute of Nuclear and Particle Physics, National Center for Scientific 
Research Demokritos, 15310 Athens, Greece}

\emailAdd{georgiou@inp.demokritos.gr}
\emailAdd{zoakos@gmail.com}

\abstract{We calculate, for the first time, three-point correlation functions involving "heavy" operators in the 
Schrodinger/null-dipole CFT correspondence at strong coupling. In particular, we focus on the three-point functions 
of the dilaton modes and two "heavy" operators. The heavy states are  dual to the single spin and dyonic magnon, 
the single spin and dyonic spike solutions or to two novel string solutions which do not have an undeformed counterpart. 
Our results provide the leading term of the correlators in the large $\lambda$ expansion and are in perfect agreement 
with the form of the correlator dictated by non-relativistic conformal invariance. 
We also specify the scaling function which can not be fixed by using conformal invariance.}

\begin{document}
\maketitle
\flushbottom


\section{Introduction and Conclusions}

Any conformal field theory (CFT) is characterized by two pieces of information. The first one is the set of its primary operators and their conformal dimensions which can be read from the two-point correlation functions. The second piece needed is the structure constants coefficients which specify the operator product expansion (OPE) of two primary operators. The knowledge of all two and three point correlators of primary operators is enough to determine through the OPE all higher point correlation functions. Usually, the aforementioned correlation functions are calculated as a series of one or more parameters, the couplings of the theory, that is they are calculated order by order in perturbation theory. It is a rare occasion when one is able to calculate the observables of the theory at large values of the coupling constants or as an exact function of the couplings. One such occasion is that of the maximally supersymmetric gauge theory in four dimensions, 
$\mathcal {N}=4$ Super Yang-Mills (SYM), one of the most thoroughly studied CFTs due to  its duality with type IIB string theory on $AdS_5 \times S^5$ \cite{Maldacena:1997re}. Exploiting the integrable structure of the theory an intense activity took place culminating in the determination of its planar spectrum for any value of the 't Hooft coupling $\lambda$. A variety of integrability based techniques were used to this end, from the asymptotic Bethe ansatz to the Y-system and the quantum spectral curve (for a review on these techniques see \cite{Beisert:2010jr}).  

Much less is known about the structure constants of the theory. The problem of determining them is much harder than that of the spectrum since the exact form of the eigenstates of the dilatation operator is also needed \cite{Georgiou:2008vk,Georgiou:2009tp,Georgiou:2011xj}.  
Systematic studies of the structure constants involving non-protected operators  were performed in \cite{Okuyama:2004bd,Roiban:2004va,Alday:2005nd,Georgiou:2012zj} by computing the corrections arising from the planar one-loop Feynman diagrams and taking into account the correct form of the one-loop eigenstates \cite{Georgiou:2012zj,Georgiou:2009tp}. 
Alternatively, one may use the string theory side in order to extract information about non-protected OPE coefficients. However this is intricate  because the supergravity limit can not be used since all non-protected operators acquire large anomalous dimensions and decouple. A particularly interesting limit in which one can extract information about  structure constants involving non-BPS operators is the BMN limit \cite{Berenstein:2002jq}. In that limit one focuses on operators with large $R$-charge which are dual to string states propagating in the PP-wave limit of $AdS_5 \times S^5$.
Different proposals for the cubic string Hamiltonian had been put forward  in \cite{Spradlin:2002ar,Pankiewicz:2002tg,DiVecchia:2003yp}.
Finally, the issue of how to correctly compare string amplitudes obtained from  the PP-wave cubic Hamiltonian and the structure constants of the $\mathcal {N}=4$ SYM was settled in \cite{Dobashi:2004nm,Lee:2004cq} by combining a number of results available from both  string and field theory \cite{Georgiou:2004ty,Georgiou:2003kt,Georgiou:2003aa,Chu:2002pd}. More recently, several integrability-based, non-perturbative in nature, methods for bootstrapping three-point correlators were developed \cite{Escobedo:2010xs,Jiang:2014mja,Basso:2015zoa,Kazama:2016cfl}. In addition, the leading term in the strong coupling expansion of three-point correlators involving three {\it heavy} states in the $SU(2)$ and  $SL(2)$ subsectors was obtained in  \cite{Kazama:2013qsa,Kazama:2011cp} by calculating the area of the corresponding minimal surface through Pohlmeyer reduction. Another interesting case is the one where the three-point correlator involves two non-protected operators dual to classical string solutions and a {\it light} state. In this case the strong coupling result  for the three-point function takes the form of the vertex operator of the {\it light} state integrated over the classical surface describing the free propagation of the {\it heavy} state from one point on the boundary of $AdS_5$ to another \cite{Zarembo:2010rr,Costa:2010rz,Roiban:2010fe,Georgiou:2010an,Bajnok:2016xxu}.

Recently, the identification of integrable deformations of the original AdS/CFT scenario has attracted a lot of attention. In these attempts it is usually supersymmetry that is partially or completely broken. However, there are cases where the deformation is more radical. One such case is the correspondence between  a certain Schrodinger spacetime and the null-dipole deformed conformal field theory \cite{Maldacena:2008wh}. The theory on the gravity side was firstly derived in \cite{Alishahiha:2003ru}. It is a solution of the type IIB equations of motion and can be obtained from the $AdS_5 \times S^5$ background through a solution generating technique known as T-s-T transformation. One performs an Abelian T-duality along one of the isometries of the sphere  $S^5$ followed by a shift along one of the light-like directions of the $AdS_5$ boundary and then performing a second T-duality along the  dualized coordinate of the sphere. The holographic dual is non-supersymmetric and realizes the Schrodinger symmetry algebra as its isometry group. The field theory dual can be obtained by introducing in the $\mathcal {N}=4$ SYM Lagrangian the appropriate $\star$-product  which can be identified with the corresponding Drinfeld-Reshetikhin twist of the underlying integrable structure of the undeformed theory \cite{Beisert:2005if,Ahn:2010ws}. As a result, it is believed that the deformed theory is fully integrable with its integrability properties inherited from the parent $\mathcal {N}=4$ SYM. 

Unlike the original AdS/CFT scenario very few observables have been computed in the deformed version of the correspondence. In \cite{Fuertes:2009ex,Volovich:2009yh} the two, three and n-point correlation functions of scalar operators were calculated using the gravity side of the correspondence. We should stress that all these operators correspond to point-like strings propagating in the $Sch_5 \times S^5$ background. Extended dyonic giant magnon and spike solutions were found in
\cite{Georgiou:2017pvi}. Their existence is in agreement with the fact that the theory is integrable. In the same work an exact in the coupling $\lambda$ expression for the dimensions of the corresponding gauge operators was
conjectured\footnote{Giant-magnon like solutions with a different dispersion relation to that of \cite{Georgiou:2017pvi} 
were studied in \cite{Ahn:2017bio}.}.  Furthermore, in the large $J$ limit agreement of this expression with the one-loop anomalous dimension of BMN-like operators was found providing further evidence in favor of the correspondence. 
On the field theory side, it is only the one-loop spectrum of operators belonging in a $SL(2)$ subsector that has been studied \cite{Guica:2017mtd} finding agreement with the string theory prediction for the anomalous dimension of certain long operators (see also \cite{Ouyang:2017yko}).  No higher point correlation functions have been calculated\footnote{Here we refer to calculations performed in the null-dipole CFT whose Lagrangian is obtained by deforming the $\mathcal {N}=4$ SYM Lagrangian with the appropriate $\star$-product.} .

In the present paper we will use the Schrodinger background in order to calculate, holographically, 
three-point functions involving two {\it heavy} operators and a {\it light} one. The  {\it light} operator will be chosen to 
be one of the modes of the dilaton while the {\it heavy} states will be either 
generalizations of the single spin and dyonic magnon or the single spin and dyonic spike solutions constructed in \cite{Georgiou:2017pvi}. We will also calculate the three point function in the case where 
the  {\it heavy} operators are two novel string solutions presented later in this work. As mentioned above the existing 
results in the literature are at the level of supergravity, in the sense that the corresponding states participating in the 
correlator are point-like and reduce to BPS states in the limit of zero deformation, i.e.  $\mu = 0$. In contradistinction, 
the {\it heavy} states we will be using are extended string solutions which tunnel from one point boundary of $Sch_5$ 
to another. These are precisely the points where the dual field theory operators will be situated. 
Our results provide the leading term of the correlators in the large $\lambda$ expansion and are in perfect agreement 
with the form of the correlator dictated by non-relativistic conformal invariance. Unlike the three-point functions in 
$\mathcal {N}=4$ SYM conformal invariance of the non-relativistic theory is not enough to fix completely the space-time 
dependence of the correlator. Instead one is left with an undetermined function, called the scaling function, of the single 
conformally invariant variable that can be built from three space-time points. 
Our calculation will determine this function $\tilde{F}(v_{12}) $ at strong coupling. 

A number of extensions and generalizations of the present calculations are possible. One may try to replace the dilaton with 
another {\it light} operator such as the $R$-current or the energy-momentum tensor $T_{\mu\nu}$. The analogous 
calculation in the case of the original AdS/CFT scenario can be found in  \cite{Georgiou:2013ff}. In order to be able to 
perform such a calculation one should first find the corresponding bulk-to-boundary propagators in the Schrodinger 
background. Another possibility would be to focus on the field theory side and try to identify the field theory operators which 
are dual to the string solutions we are using in this work and calculate their correlation functions at weak coupling possibly employing the 
integrable structure of the theory. In particular, it would be interesting to focus on operators which can be described by 
coherent states since in this case one may be able to compare the weak and strong coupling results along the lines of 
\cite{Escobedo:2011xw,Georgiou:2011qk}. Finally, it would be interesting to employ integrability in order to calculate 
three-point correlation functions involving three {\it heavy} operators or one {\it heavy} and two  {\it light} operators
 both holographically and at weak coupling.


\section{Classical string solutions}

In this section we will write down and analyze the four classical string solutions, which in the following section will
be used in the calculation of the three-point correlation functions. 
In subsection \ref{GM-SS} we review the dyonic magnon and dyonic spike solutions, in \ref{GM} we review the single spin giant magnon and in \ref{SS} we review the single spin spike solutions. These solutions were initially presented and studied in  \cite{Georgiou:2017pvi}. To be precise, we will consider slight generalizations of the string solutions presented in  \cite{Georgiou:2017pvi} because these solutions will have a clear interpretation as strings tunneling from one point of the boundary to another and as a result will be of immediate use in the calculation of the three-point correlators. 
Subsequently, in subsections \ref{SpinBMN} and \ref{New-Classical-solution} 
we present two new classical string solutions. 
The solution of \ref{SpinBMN} is a generalization of the spinning BMN-like string solution that first appeared in  
\cite{Guica:2017mtd}. In \ref{New-Classical-solution} we present a completely new solution, 
with an oscillating behavior along the holographic 
direction 
which does not have an undeformed analogue. 

We consider the following consistent truncation
of the 10d $Sch_5\times S^5$ metric\footnote{With respect to the notation of \cite{Georgiou:2017pvi}
for the $S^3$ we have performed the following change of variables 
\begin{equation} \label{S3coord}
\theta\,=\,2\,\eta
\hspace{1.5cm}
\psi\,=\,\varphi_1+\,\varphi_2
\hspace{1.5cm}
\phi\,=\,\varphi_1-\,\varphi_2 \, . 
\end{equation} 
More details about the consistent truncation can be found in appendix \ref{Polyakov-truncation}. }
\begin{eqnarray} \label{metric}
ds^2 & = & - \left(1\, + \, \frac{\mu^2}{Z^4} \right) dT^2 \,  + \,  
\frac{1}{Z^2} \, \Bigg(2 dT dV + dZ^2 + d{\vec X}^2 - {\vec X}^2  dT^2 \Bigg)  \, + \, ds^2_{S^3}
\quad {\rm with} 
\nonumber \\ [7pt]
ds^2_{S^3} & = & d\eta^2 \, + \, \sin^2\eta \, d\varphi_1^2 + \cos^2\eta \, d\varphi_2^2 
\end{eqnarray}
that is supplemented by the following B-field
\begin{equation} \label{Bfield}
B \, = \,  \frac{\mu}{Z^2} \, d T \wedge \Big( \sin^2\eta \, d\varphi_1 + \cos^2\eta \, d\varphi_2 \Big) \, . 
\end{equation}


\subsection{Dyonic giant magnon and dyonic single spike}
\label{GM-SS}

Here we review and extend the dyonic giant magnon and single spike solutions that were originally presented in 
\cite{Georgiou:2017pvi}\footnote{The solutions of \cite{Georgiou:2017pvi}  have ${\vec X}_0=0$.}, since we intend to use them in the three-point function calculation.
We consider the following ansatz, for both solutions, 
\begin{eqnarray} \label{ansatz-GM-SS}
&& 
T\,=\, \kappa \,\tau \, , 
\quad
V \,=\,\alpha \, \tau \,  - \, \frac{\vec X^2_0}{4} \, \sin 2 \kappa  \tau \, + \, V_y(y)\, , 
\quad 
{\vec X} \,=\, {\vec X}_0 \, \sin \kappa  \tau \, ,
\nonumber \\ [5pt]
&&
Z \,=\, Z_0 \, , 
\quad 
\eta\,=\, \frac{1}{2} \,\theta_y(y) \, , 
\quad
\varphi_{1/2}=\frac{1}{2} \left(\omega_{\psi} \pm \omega_{\phi} \right)\tau+ \frac{1}{2}\Big(\Psi_y(y) \pm \Phi_y(y)\Big) \, , 
\end{eqnarray} 
where we have defined the variable $y$ as  
\begin{equation} \label{definition-y}
y\, \equiv \, c\,\sigma-d\,\tau \, . 
\end{equation}
The explicit expressions for the functions $V_y(y)$, $\theta_y(y)$, $\Psi_y(y)$ and $\Phi_y(y)$ 
as well as for the constants $\kappa$ and $Z_0$ can be found in \cite{Georgiou:2017pvi} and 
it is where the interested reader is referred to.  The new dispersion relation for the giant magnon reads
\begin{equation}
\label{dispersionGM-extended}
\sqrt{\left(E \, - \, \frac{\vec X_0^2}{2} \, M\right)^2 \, - \, \mu^2 \, M^2} \, - \, J_1\,=\,
\sqrt{J_2^2 \, + \, \frac{\lambda}{\pi^2} \, \sin^2\frac{p}{2}}
\end{equation}
while for the single spike 
\begin{equation}
\label{dispersionSS-extended}
J_1^2 \, - \, J_2^2 
\,=\, \frac{\lambda}{\pi^2} \, 
\sin^2\left[\frac{ \pi}{\sqrt{\lambda}} \, \left(E \, - \frac{\vec X_0^2}{2} \, 
M \, - \, \mu \, M \right) - \, \frac{1}{2} \, \Delta \varphi \right]  \, .
\end{equation}
Setting the constant ${\vec X_0}$ to zero, i.e. ${\vec X_0}=0$, we obtain the dispersion relations of 
\cite{Georgiou:2017pvi}.

In order to compute the three-point correlation function we need to Euclideanize the world-sheet metric 
and then rewrite the solution \eqref{ansatz-GM-SS}  in Poincare coordinates. 
To this end we have used the coordinate transformation relating the global to the Poincare coordinates
which can be found in \cite{Blau:2009gd}. Namely,
\begin{equation} \label{transf}
x^+ =\tan T \, , \qquad 
x^-=V-\frac{1}{2} \left(Z^2+\vec{X}^2\right) \, \tan T \, , \qquad 
z=\frac{Z}{\cos T} \, ,\qquad 
\vec{x}=\frac{\vec{X}}{\cos T}
\end{equation}
where $(x^+,x^-,z,\vec{x})$ parametrize the space in Poincare coordinates.
Performing these two steps we arrive at the following solution
\begin{eqnarray} \label{ansatz-GM-SS-Poincare}
&&
\mathbf{x} \, = \, \frac{\mathbf{X}}{2}\,\tanh\kappa \tau \, , 
\hspace{1cm}
x^+ \, = \, i \, \frac{\mathbf T}{2}\,\tanh\kappa \tau \, , 
\hspace{1cm}
z \, = \, \sqrt{\frac{\mathbf T}{2}}\, \frac{Z_0}{\cosh\kappa \tau }\, ,
\nonumber \\[5pt]
&& x^- \, = \, - \, \frac{i}{2} \, \left[Z_0^2 \, - \, 
\frac{\mathbf{X}^2}{2\, \mathbf T}\right]\,\tanh\kappa \tau \, + \, i \, \alpha \, \tau  \, + \, V_y(y) \, ,
\\ [5pt]
&&
\eta\,=\, \frac{1}{2} \,\theta_y(y)\, ,
\hspace{1cm}
\varphi_{1/2}=\frac{i}{2} \left(\omega_{\psi}\pm\omega_{\phi} \right)\tau+ \frac{1}{2}\left(\Psi_y(y)\pm\Phi_y(y)\right)
\nonumber
\end{eqnarray}
where the definition for $y$ becomes  
\begin{equation} \label{definition-y-poincare}
y\, \equiv \, c\,\sigma- i \, d\,\tau
\end{equation}
due to the Wick rotation performed to the world-sheet time. To obtain \eqref{ansatz-GM-SS-Poincare} we have also
redefined ${\vec X_0}$ as ${\vec X_0} = - \frac{i}{2} \,  {\mathbf X} $. 
The complex solution \eqref{ansatz-GM-SS-Poincare} describes a string propagating from the point 
$\mathbf{x}(\tau=-\infty)=- {\mathbf{X} \over 2}$ and $x^+(\tau=-\infty)=- i \, \frac{\mathbf T}{2}$ of the boundary -which is located at $z=0$- to another point of the boundary with coordinates $\mathbf{x}(\tau=\infty)= {\mathbf{X} \over 2}$ and $x^+(\tau=\infty)=i \, \frac{\mathbf T}{2}$. The two {\it heavy}  field theory operators of the dual CFT will be inserted at precisely the aforementioned points.

In what follows we will also need the Lagrangian densities evaluated on the classical solutions. These will be useful in the calculation of the corresponding three-point functions. Computing the Lagrangians on-shell we obtain 
the following expressions\footnote{For the giant magnon solution and for the new classical string solution 
we use the notation $d=v \,c$. For the single spike solution we use the notation $c=v \,d$. In all cases $0<v<1$.}
\begin{equation} \label{Lag-onshell-GM-2spin}
{\cal L}_{\text{on shell}}^{\text{GM 2s}} \, =  \, - \, i \, \frac{\sqrt{\lambda}}{2 \, \pi}\,
\frac{\omega_{\phi}\, \omega_{\psi} }{1 \, - \, v^2} \, \left(1+\frac{\alpha \, v}{ \mu \,  \omega_{\psi}}\right)\, u(y)
\end{equation}
and
\begin{equation} \label{Lag-onshell-SS-2spin}
{\cal L}_{\text{on shell}}^{\text{SS 2s}} \, =  \, - \, i \, \frac{\sqrt{\lambda}}{2 \, \pi}\,
\frac{\alpha\, v \,\left(\omega_{\psi} + \omega_{\phi}\right)}{2 \, \mu} \, \Bigg[1- \frac{v \, \omega_{\phi}}{1 - v^2} \,
\frac{2 \, \alpha - v \, \mu \, \left(\omega_{\phi} -\omega_{\psi} \right)}{\alpha \, v \, \left(\omega_{\psi} + 
\omega_{\phi}\right)} \, u(y) \Bigg]
\end{equation}
where the explicit expression of the function $u(y)$, both for the giant magnon and the spike solutions, can be found 
in \cite{Georgiou:2017pvi}.


\subsection{Single spin giant magnon}
\label{GM}

Here we review and extend the single spin giant magnon solution that can also be obtained from the dyonic one, 
once we set the conserved momentum $J_{2}$ equal to zero, i.e. $J_{2} = 0$. 
The ansatz we consider is the following\footnote{In all the expressions (for the functions and the constants) 
we present in the current and in the following subsections, the 
boundary conditions have been taken into account.} 
\begin{eqnarray} \label{ansatz-GM}
&& 
T\,=\, \kappa \,\tau \, , 
\quad
V \,=\,\mu^2 \, m \, \tau -  \frac{\vec X^2_0}{4} \, \sin 2 \kappa \, \tau \,  + \, V_y(y) \, , 
\quad
{\vec X} \,=\, {\vec X}_0 \, \sin \kappa  \tau \, ,
\nonumber \\ [5pt]
&& 
Z \,=\, \sqrt{\frac{\kappa}{m}} \, , 
\quad
\eta\,=\, \frac{1}{2} \,\theta_y(y) \, ,
\quad
\varphi_1\,=\,\omega \,\tau\,- \, \mu \, m\,\sigma \,+ \,\Phi_y(y) \, ,
\quad
\varphi_2\,=\,- \, \mu \, m \,\sigma 
\end{eqnarray} 
where $y$ is defined in \eqref{definition-y}.   
As in the dyonic case \cite{Georgiou:2017pvi}, we rewrite the differential equation for $\theta_y(y)$ in terms of a new function 
 $u(y) \equiv \cos^2 \left(\frac{1}{2}\, \theta_y\right)$ as 
\begin{equation} \label{equation-u} 
\frac{(u')^2}{2}\,+\,{\cal W}(u) \,=\,0
\hspace{.7cm} \textrm{with} \hspace{.7cm}
{\cal W}(u)\,=\,-\,2 \left(\beta_6\, u + \beta_4 \right) \, u^2
\end{equation}
where the two constants $\beta_6$ and $\beta_4$ have the values
\begin{eqnarray}
\label{beta6GM}
\hspace{-.7cm}
\beta_6 \,& = & \, - \,  \frac{c^2 \, \omega^2}{(c^2-d^2)^2} \, 
= \, - \, \frac{\omega^2}{c^2 \left(1\, - \, v^2\right)^2}  \, < 0
\\
\label{beta4GM}
\hspace{-.7cm}
\beta_4 \,& = & \, \frac{\left(\mu\, m \, d - \, c \, \omega \right)^2 \, - \, c^2 \, \mu^2 \, m^2}{c^2
   \left(c^2-d^2\right)} \, = \, \frac{\left(\mu \, m \, v \, - \, \omega \right)^2 \, - \, 
   \mu^2 \, m^2}{c^2 \, \left(1\, - \, v^2\right)} \, 
\end{eqnarray}
and the prime denotes differentiation with respect to $y$.
In order for $\beta_4$ to be positive\footnote{Otherwise the coordinate $\theta(y)$ will have oscillatory behavior as 
$y\rightarrow \pm \infty$ and this is not the expected behavior for the giant magnon.} 
we constrain the values of $\omega$ as follows 
\begin{equation}
\omega < - \mu \, m \, \left(1-v \right) 
\quad {\rm or} \quad
\omega > \mu \, m  \, \left(1+ v \right) \, . 
\end{equation}
The solution of equation \eqref{equation-u} is known from the analysis in \cite{Georgiou:2017pvi} and is given by 
\begin{equation}
\label{u-solution}
u(y)\,=\, \frac{\beta_4}{|\beta_6|}\, \frac{1}{\cosh^2\big(\sqrt{\beta_4}\, y\big)} \, . 
\end{equation}
The equations of motion give the following expressions for $V_y'(y)$ and $\Phi_y'(y)$ 
\begin{equation} \label{constraintVGM}
V_y'(y) \, = \, - \,  \frac{\mu \, \omega}{c \, \left(1\, - \, v^2\right)}  \, u(y)
\end{equation}
and 
\begin{equation} \label{constraintPhiGM}
\Phi_y'(y) \, = \, \frac{\mu\, m}{c} \, +  
\frac{\mu \, m  \left(1 \, - \, v^2\right) \, + \,  v \, \omega }{c \,  \left(1\, - \, v^2\right)} \, \frac{u(y)}{1-u(y)} \, . 
\end{equation}
The value of $\kappa$ is determined by the Virasoro constraint $G_{\mu \nu} (\dot{X}^{\mu}\dot{X}^{\nu}+X'^{\mu} X'^{\nu})=0$ as follows
\begin{equation}
\kappa^2\, = \,\mu ^2 \,m^2 \left(1+ v^2\right)\, - \, 2 \,\mu \, m \,v \,\omega + \omega ^2 \, . 
\end{equation}
As a result the dispersion relation of the solutions becomes
\begin{equation}
\label{dispersionGM}
\sqrt{\left(E \, - \, \frac{\vec X_0^2}{2} \, M\right)^2  \, - \, \mu^2 \, M^2} \, - \, J \,=\,
\frac{\sqrt{\lambda}}{\pi} \sin\,  \frac{p}{2} \, . 
\end{equation}
This dispersion relation can be derived, as usual, by finding the relation among finite combinations of 
the infinite conserved charges $E,\,J \,{\rm and} \,M$.
Euclideanizing \eqref{ansatz-GM} 
and rewriting it in Poincare coordinates we obtain that
\begin{eqnarray} \label{ansatz-GM-Poincare}
&&
\mathbf{x} \, = \, \frac{\mathbf{X}}{2}\,\tanh\kappa \tau \, , 
\quad 
x^+ \, = \, i \, \frac{\mathbf T}{2}\,\tanh\kappa \tau \, , 
\quad 
z \, = \, \sqrt{\frac{\kappa \, \mathbf T}{2 \, m}}\, \frac{1}{\cosh\kappa \tau }\, ,
\nonumber \\[5pt]
&& x^- \, = \, - \, \frac{i}{2} \, \left[\frac{\kappa}{m} \, - \, 
\frac{\mathbf{X}^2}{2\, \mathbf T}\right]\,\tanh\kappa \tau \, + \, i \, \mu^2 \, m \, \tau  \, + \, V_y(y) \, ,
\nonumber \\ [5pt]
&& \eta\,=\, \frac{1}{2} \,\theta_y(y)\, , 
\quad
\varphi_1\,=\, i \, \omega\,\tau\,- \, \mu \, m\,\sigma \,+ \,\Phi_y(y)
\quad \& \quad
\varphi_2\,=\,- \, \mu \, m \,\sigma \, , 
\end{eqnarray}
where now $y$ is defined in \eqref{definition-y-poincare}. 
Computing the Lagrangian on shell it is straightforward to obtain
\begin{equation} \label{Lag-onshell-GM}
{\cal L}_{\text{on shell}}^{\text{GM 1s}} \, =  \, - \, i \, \frac{\sqrt{\lambda}}{2 \, \pi}\,
\frac{\omega^2}{1 \, - \, v^2} \, \left(1-\frac{\mu \, m \, v}{ \,  \omega}\right)\, u(y).
\end{equation}
Notice that setting $\omega_{\psi}=\omega_{\phi} -2 \, \mu \, m \, v$ with $\omega_{\phi}  \equiv \omega$ and 
$\alpha =\mu^2 \, m $ in \eqref{Lag-onshell-GM-2spin} the on-shell Lagrangian of the dyonic magnon becomes that of the single spin giant magnon \eqref{Lag-onshell-GM}.


\subsection{Single spin single spike}
\label{SS}

Here we review and extend the single spin single spike solution. 
This solution can be found by considering the following ansatz 
\begin{eqnarray} \label{ansatz-SS}
&& 
T\,=\, \kappa \,\tau \, , 
\quad
V \,=\,\alpha \, \tau \,  - \, \frac{\vec X^2_0}{4} \, \sin 2 \kappa  \tau \, + \, V_y(y)\, , 
\quad 
{\vec X} \,=\, {\vec X}_0 \, \sin \kappa  \tau \, ,
\nonumber \\ [5pt]
&& 
Z \,=\, \sqrt{\frac{2 \, d \, \kappa \,  \mu^2 }{d \, \alpha -c \, \omega \, \mu}} \, , 
\quad
\eta\,=\, \frac{1}{2} \,\theta_y(y)\, , 
\nonumber \\ [5pt]
&&
\varphi_1\,=\,\omega \,\tau\,+ \frac{1}{2} \left(\frac{\alpha}{\mu}- \, \frac{c \, \omega }{d} \right)\,
\sigma \,+ \,\Phi_y(y)
\quad \& \quad
\varphi_2\,=\,- \, \frac{1}{2} \left(\frac{\alpha}{\mu}- \, \frac{c \, \omega }{d} \right) \,
\sigma  \, , 
\end{eqnarray} 
where $y$ is  defined again by \eqref{definition-y}.   
After some algebra the equations of motion lead to a differential equation analogous to \eqref{equation-u} but with the constants taking the values
\begin{eqnarray}
\label{beta6SS}
\hspace{-.7cm}
\beta_6 \,& = & \, - \,  \frac{\alpha^2 \, d^2}{\mu^2 \, \left(c^2-d^2\right)^2} \, = \, -\frac{\alpha^2}{\mu^2 \, 
d^2 \left(1\, - \, v^2\right)^2} 
\, < 0
\\
\label{beta4SS}
\hspace{-.7cm}
\beta_4 \,& = & \, \frac{\alpha^2}{\mu^2 \,\left(d^2-c^2\right)} \, = \, \frac{\alpha^2}{\mu^2\, d^2 \left(1\, - \, v^2\right)}\, > 0  \, . 
\end{eqnarray}
Since $\beta_4 > 0$ there is no constraint on the values of $\omega$, contrary to the giant magnon case, and the 
solution for $u(y)$ is given by \eqref{u-solution}, with $\beta_6$ and $\beta_4$ defined in equations \eqref{beta6SS}
and  \eqref{beta4SS} respectively.
The expressions for $V_y'(y)$ and $\Phi_y'(y)$ take the form
\begin{equation} \label{constraintVSS}
V_y'(y) \, = \, \frac{\alpha}{d \, \left(1\, - \, v^2\right)}  \, u(y)
\end{equation}
and 
\begin{equation} \label{constraintPhiSS}
\Phi_y'(y) \, = \, \frac{\omega}{d} \, - \,  
\frac{\alpha  \, v}{\mu \, d \,  \left(1\, - \, v^2\right)} \, \frac{u(y)}{1-u(y)} \, . 
\end{equation}
The dispersion relation of the single spin spike becomes
\begin{equation}
\label{dispersionSS}
J^2 \,=\, \frac{\lambda}{\pi^2} \, 
\sin^2\left[\frac{ \pi}{\sqrt{\lambda}} \, \left(E \, - \,  \frac{\vec X_0^2}{2} \, 
M \,- \, \mu \, M \right) - \, \frac{1}{2} \, \Delta \varphi \right]  \, .
\end{equation}
As we did in the giant magnon cases the next step consists of euclideanizing the world-sheet time of the solution \eqref{ansatz-SS} 
and rewriting it in Poincare coordinates to get
\begin{eqnarray} \label{ansatz-SS-Poincare}
&&
\mathbf{x} \, = \, \frac{\mathbf{X}}{2}\,\tanh\kappa \tau \, , 
\quad 
x^+ \, = \, i \, \frac{\mathbf T}{2}\,\tanh\kappa \tau \, , 
\quad 
z \, = \, \sqrt{\frac{d \, \kappa \, \mu^2 \, \mathbf T}{\alpha \, d \, - \, c \, \mu \, \omega }}\, \frac{1}{\cosh\kappa \tau }\, .
\nonumber \\[5pt]
&& x^- \, = \, - \, \frac{i}{2} \, \left[\frac{2 \, d \, \kappa \, \mu ^2}{\alpha \, d \, - \, c \, \mu  \omega } \, - \, 
\frac{\mathbf{X}^2}{2\, \mathbf T}\right]\,\tanh\kappa \tau \, + \, i \, \alpha \, \tau  \, + \, V_y(y) \, ,
\quad 
\eta\,=\, \frac{1}{2} \,\theta_y(y) \, , 
\nonumber \\ [5pt]
&&
\varphi_1\,=\, i \, \omega \,\tau\,+ \frac{1}{2} \left(\frac{\alpha}{\mu}- \, \frac{c \, \omega }{d} \right)\,
\sigma \,+ \,\Phi_y(y) \, , 
\quad \& \quad 
\varphi_2\,=\,- \, \frac{1}{2} \left(\frac{\alpha}{\mu}- \, \frac{c \, \omega }{d} \right) \,
\sigma 
\end{eqnarray} 
where now $y$ is defined in \eqref{definition-y-poincare}. 
Computing the Lagrangian on shell we have
\begin{equation} \label{Lag-onshell-SS}
{\cal L}_{\text{on shell}}^{\text{SS 1s}} \, =  \, - \, i \, \frac{\sqrt{\lambda}}{2 \, \pi}\,
\frac{1}{2} \, \frac{\alpha^2}{\mu^2} \, \left(1+ \frac{\mu \, v \, \omega}{\alpha} \right) \, \left(1-\, \frac{u(y)}{1-v^2}\right) \, .
\end{equation}
Notice that setting $\omega_{\phi}= \frac{\alpha}{\mu \, v }$ with $\omega_{\psi}  \equiv \omega$, 
the on-shell Lagrangians of the dyonic and single spin spikes, 
namely \eqref{Lag-onshell-SS-2spin} and \eqref{Lag-onshell-SS}, are identified.


\subsection{Spinning BMN-like strings}
\label{SpinBMN}

In this section, we will present a solution generalizing the BMN-like solution of \cite{Guica:2017mtd}. 
Our solution is also winding one of the isometries of the $S^5$ with the winding number being $n \in \mathbb{Z}$. 
In the limit of zero winding $n = 0$ and $\vec X_0=0$ we obtain the spinning BMN-like solution that was presented in \cite{Guica:2017mtd}. 

To start we consider the following ansatz for the spinning BMN-like solutions
\begin{eqnarray} \label{ansatz-spinning}
&& 
T\,=\, \kappa \,\tau \, , 
\quad
V \,=\,\mu^2 \, m \, \tau -  \frac{\vec X^2_0}{4} \, \sin 2 \kappa \, \tau \,  - 
\, \frac{\omega \,  n \, \sigma}{m\,\left(1 \, - \, 
\frac{n}{\mu \, m }\right)}  \, , 
\quad
{\vec X} \,=\, {\vec X}_0 \, \sin \kappa  \tau \, ,
\nonumber \\ [5pt]
&& 
Z \,=\, \sqrt{\frac{\kappa}{m}} \, \, \frac{1}{\sqrt{\left(1 \, - \, \frac{n}{\mu \, m }\right)}} \, ,
\quad
\eta\,=\,\frac{\pi}{2} \, , 
\quad
\varphi_1\,=\,\omega \, \tau\, + \, n \,\sigma
\quad \& \quad 
\varphi_2\,=\, 0
\, . 
\end{eqnarray} 
In order for $Z$, in \eqref{ansatz-spinning}, to be real the parameter $n$ should satisfy the constraint $n< \mu \, m$.
The $G_{\mu \nu} (\dot{X}^{\mu}\dot{X}^{\nu}+X'^{\mu} X'^{\nu})=0$ Virasoro constraint requires that
\begin{equation}  \label{virasoro-spinning}
\kappa^2 \, = \,  \omega^2 \, + \, \mu^2 \, m^2 
\end{equation}
while the parameters of the solution are related to the conserved charges as follows 
\begin{equation} \label{conserved-spinning}
E \, - \,  \frac{\vec X_0^2}{2} \, M \, = \,  \sqrt{\lambda} \, \kappa \, ,  
\quad 
J \, = \, \sqrt{\lambda} \, \omega  
\quad \& \quad 
M \, = \,  \sqrt{\lambda}  \,  \left(m \, - \, \frac{n}{\mu }\right) \, . 
\end{equation}
As a result the dispersion relation is given by the following expression
\begin{equation} \label{dispersion-spining}
\Bigg(E\, - \,  \frac{\vec X_0^2}{2} \, M \Bigg)^2  \, = \, J^2 \, + \,  
\mu^2 \, \Bigg( M \, + \, \sqrt{\lambda} \, \, \frac{n}{ \mu} \Bigg)^2 \, . 
\end{equation}
Euclideanizing \eqref{ansatz-spinning} 
and rewriting it in Poincare coordinates we have
\begin{eqnarray} \label{ansatz-spinning-Poincare}
&&
\mathbf{x} \, = \, \frac{\mathbf{X}}{2}\,\tanh\kappa \tau \, , 
\quad 
x^+ \, = \, i \, \frac{\mathbf T}{2}\,\tanh\kappa \tau \, , 
\nonumber \\[5pt]
&& x^- \, = \, - \, \frac{i}{2} \, \left[\frac{\kappa }{m}\, \left(1 \, - \, \frac{n}{\mu \, m }\right)^{-1} \, - \, 
\frac{\mathbf{X}^2}{2\, \mathbf T}\right]\,\tanh\kappa \tau \, + \, i \, \mu^2 \, m \, \tau  - 
\, \frac{\omega \,  n \,  \sigma}{m\,\left(1 \, - \, \frac{n}{\mu \, m }\right)} \, ,
\nonumber \\[5pt]
&&
\varphi_1\,=\, i \, \omega \,\tau\, + \, n \,\sigma \, ,
\quad 
z \, = \, \sqrt{\frac{\kappa \, \mathbf T}{2 \, m}\, \left(1 \, - \, \frac{n}{\mu \, m }\right)^{-1} }\,
\frac{1}{\cosh\kappa \tau }\, .
\end{eqnarray}
Subsequently. we compute the Lagrangian on shell, to obtain the following result
\begin{equation} \label{Lag-onshell-spinning}
{\cal L}_{\text{on shell}}^{\text{BMN}} \, = \, - \, i \, \frac{\sqrt{\lambda}}{2 \, \pi}\, \mu \, m \, n \, .
\end{equation}
Note that for the BMN-like solution of \cite{Guica:2017mtd} $n=0$ and as a result the on shell Lagrangian becomes zero.


\subsection{New classical string solution}
\label{New-Classical-solution}

In this subsection we will describe a new classical string solution that does not have an undeformed analog. 
We start by writing the ansatz we will use and is inspired by the giant magnon solution of \cite{Georgiou:2017pvi}, with the 
novelty that now the function $Z$ is not a constant but a function of the worldvolume coordinates, that is
\begin{eqnarray} \label{ansatz-NS}
&&
T\,=\, \kappa \,\tau \, + \, T_y(y) \, , 
\quad
V \,=\,\mu^2 \, m \, \tau \,  - \, \frac{\vec X^2_0}{4} \, \sin 2 \Big(\kappa  \tau + \, T_y(y) \Big)\, + \, V_y(y) \, , 
\quad 
Z \,=\, \sqrt{\frac{\kappa}{m}} \, Z_y(y) \, , 
\nonumber \\ [5pt]
&&
{\vec X} \,=\, {\vec X}_0 \, \sin \Big(\kappa  \tau + \, T_y(y) \Big) \, ,
\quad
\eta\,=\,\frac{\pi}{2} \, , 
\quad
\varphi_1\,=\,\omega \,\tau\,+\,\Phi_y(y)
\quad \& \quad 
\varphi_2\,=\, 0
\end{eqnarray} 
where $y$ is defined in \eqref{definition-y}. 
Note that if we set $T_y=V_y=\Phi_y={\vec X}_0=0$ and $Z_y=1$ in the solution \eqref{ansatz-NS} we get the spinning string 
solution of \cite{Guica:2017mtd}. As usual the functions $T_y(y)$, $V_y(y)$, $Z_y(y)$ and $\Phi_y(y)$ 
are functions of $\sigma$ and $\tau$ and they will be determined through the equations of motion and the 
Virasoro constraints. 

Following the same line of reasoning as in \cite{Georgiou:2017pvi} and fixing the integration constants in such a way that 
if we set $Z_y = 1$ we obtain the spinning string solution of \cite{Guica:2017mtd},
we have the following first order differential equations for the functions $T'_y(y)$, $V'_y(y)$ and $\Phi'_y(y)$ 
\begin{equation} \label{constraintT-NS}
T'_y(y) \, = \, \frac{d \, \kappa}{c^2-d^2} \, \left(Z_y^2 \, - \, 1 \right) \, , 
\end{equation}
\begin{equation} \label{constraintV-NS}
V_y'(y) \, = \, \frac{1}{c^2-d^2} \, \left(Z_y^2 \, - \, 1 \right)\, \Bigg[\frac{d\,  \kappa ^2 \, Z_y^2}{m}\, - 
\, c\,  \mu  \, \omega \Bigg] \, , 
\end{equation}
\begin{equation} \label{constraintPhi-NS}
\Phi_y'(y) \, = \, 
\frac{c \, m \, \mu}{c^2-d^2}\, \left(1 - \, \frac{1}{Z_y^2} \right)\, .
\end{equation}
From the two Virasoro constraints, we obtain a first order differential equation for $Z'_y$ 
(which is actually consistent with the second order differential equation coming from the variation along the direction $Z$) 
\begin{equation} \label{constraintZ}
\left(Z'_y\right)^2 \, = \, \Bigg[ \frac{d \, \kappa}{c^2-d^2} \, \left(Z_y^2 \, - \, 1 \right) \Bigg]^2 \, 
\left[\frac{2 \, c \, \mu \,  m \, \omega}{d \, \kappa ^2} \, - \, Z_y^2 \right]
\end{equation}
as well as the following algebraic constraint
\begin{equation} \label{algebraic}
\kappa^2 \, = \, \mu^2 \, m^2 \, + \, \omega^2 \, . 
\end{equation}
Since the RHS of \eqref{constraintZ} should be positive, the value of $Z_y$ is limited in the 
following interval  
\begin{equation}
0 < Z_y < Z_{\rm qrit} \equiv \sqrt{\frac{2 \, c \, \mu \,  m \, \omega}{d \, \kappa ^2} } \, . 
\end{equation}
As can be seen from \eqref{constraintZ} as soon as we set the deformation parameter to zero, i.e. $\mu =0$, 
the only acceptable solution is that the LHS and the RHS of \eqref{constraintZ} vanish independently, i.e. $Z_y=1$. In that 
way we are led to the spinning BMN-like string solution of \cite{Guica:2017mtd}.

In order to solve the equation of motion for $Z_y$ \eqref{constraintZ} we need to specify whether $Z_{\rm qrit}>1$ or  
$Z_{\rm qrit}<1$. In what follows we will solve the equation in the interval 
\begin{equation} \label{rangeZy}
1 < Z_y < Z_{\rm qrit}
\end{equation}
since when $Z_{\rm qrit}<1$ the variable $y$ cannot range to $\pm \infty$.
Integrating  \eqref{constraintZ} and imposing 
\begin{eqnarray}
 y \, = \, 0  \quad & \Rightarrow &  \quad Z_y \, = \, Z_{\rm qrit} 
\nonumber \\
y \, = \, \pm \, \infty  \quad & \Rightarrow &  \quad Z_y \, = \, 1  
\end{eqnarray}
we obtain the following solution
\begin{equation}
Z_y(y) \, = \,  \frac{Z_{\rm qrit}}{\sqrt{1\, + \, \left(Z_{\rm qrit}^2-1\right) 
\tanh ^2\left[\frac{d \, \kappa \sqrt{Z_{\rm qrit}^2-1}}{c^2-d^2} \, y\right]}} \, .
\end{equation}
Plotting the above solution it is easy to verify that $Z_y$  takes the value $1$ at $y=-\infty$ reaching its maximum value $Z_{\rm qrit}$ at $y=0$ and $Z_{\rm qrit}$ and goes back to $Z_y=1$ at $y=\infty$, with $Z'_y>0$ for $y \in [-\infty,0]$ and $Z'_y<0$ for $y \in [0,+\infty]$. 

Now we can calculate the conserved charges for the new classical solution and construct the finite 
linear combinations of those conserved quantities that will lead us to the dispersion relation. 
In the following we will also use the following change of variables
\begin{equation} \label{change-variables}
d \sigma = \frac{1}{c}\, \frac{d Z_y}{|Z'_y|}
\quad \Rightarrow \quad
\int_{-\infty}^{+\infty} d \sigma = \frac{2}{c} \int_{1}^{Z_{\rm qrit}} \frac{d Z_y}{|Z'_y|} \, . 
\end{equation}
Using the above change of variables it is possible to calculate the three conserved charges 
$\mathcal{E}$,  $\mathcal{M}$ and $\mathcal{J}$\footnote{We use the notation $\mathcal{E} \equiv E/T$, 
$\mathcal{M} \equiv M/T$ and $\mathcal{J} \equiv J/T$.}. All these charges diverge and it is possible to express them in terms 
of a non-convergent integral. That integral should be eliminated  among the conserved quantities  in order to construct the dispersion relation. 
The non-convergent integral reads
\begin{equation}
{\cal I}_{\rm div} \, \equiv \, \int_{1}^{Z_{\rm qrit}} \frac{d \xi}{\left(\xi^2 \,- \,1\right) \, \sqrt{Z_{\rm qrit}^2 \, - \, \xi^2}} \, . 
\end{equation}
The conserved charges are given by
\begin{eqnarray}
\label{Energy-NS}
\mathcal{E} \, - \,  \frac{\vec X_0^2}{2} \, \mathcal{M}
&=& \frac{2\left(1\, - \, v^2\right)}{v} \, \,{\cal I}_{\rm div} \, - \, \frac{2\,m\,\mu\,\omega}{\kappa^2} \, 
\frac{\sqrt{Z_{\rm qrit}^2-1}}{Z_{\rm qrit}^2}
\\
\label{M-NS}
\rule{0pt}{.8cm}
\mathcal{M}&=& \frac{2 \, m \, \left(1\, - \, v^2\right)}{\kappa \, v} \,\, {\cal I}_{\rm div} \, - \, \frac{2\,m}{ \kappa \, v} \, 
\frac{\sqrt{Z_{\rm qrit}^2-1}}{Z_{\rm qrit}^2}
\\
\label{J-NS}
\rule{0pt}{.8cm}
\mathcal{J}
&=&  \frac{2 \, \omega \, \left(1\, - \, v^2\right)}{\kappa \, v} \,\, {\cal I}_{\rm div} \, . 
\end{eqnarray}
Using the algebraic constraint \eqref{algebraic} we eliminate the non-convergent integral and construct the following
dispersion relation
\begin{equation} \label{dispersion-NS}
\Bigg[E \, - \,  \frac{\vec X_0^2}{2} \, M \, + \, \frac{\sqrt{\lambda}}{2 \pi}\,\sqrt{\frac{2 \, \mu \, m \,\omega \, v}{\mu ^2 \, m^2 + 
\omega^2} \, - \, v^2} \Bigg]^2 -
\Bigg[\mu \, M \, + \, \frac{\sqrt{\lambda}}{2 \pi}\, \sqrt{\frac{2 \, \mu \, m}{v\, \omega } - 
1 - \frac{\mu ^2\, m^2}{\omega^2} }\Bigg]^2 = J^2 \, . 
\end{equation}
In order to write the dispersion relation in terms of conserved quantities we introduce the following 
ratio of the parameters $\mu$, $m$ and $\omega$ that we denote by $\chi$ and which repeatedly appears in \eqref{dispersion-NS}
\begin{equation}
\chi \, \equiv \, \frac{\mu \, m}{\omega} \, . 
\end{equation}
Using linear combinations of \eqref{Energy-NS}, \eqref{M-NS} and \eqref{J-NS} it is possible to 
find an expression for $\chi$  in terms of $E$, $M$ and $J$ which reads
\begin{equation} 
\frac{J\, \sqrt{1+\chi^2} - \left(E - \frac{\vec X_0^2}{2} \, M\right)}{J \, \chi - \mu \, M} \, = \, \frac{v}{\sqrt{1+\chi^2}} \, . 
\end{equation}
In terms of $\chi$, the conserved charges and the velocity $v$ of the soliton the dispersion relation becomes finally
\begin{equation} \label{dispersion-NS-v2}
\Bigg[E \, - \,  \frac{\vec X_0^2}{2} \, M \, + \, \frac{\sqrt{\lambda}}{2 \pi}\, \frac{v}{\sqrt{1+\chi^2}}\, \sqrt{\frac{2\, \chi}{v}-1-\chi^2}\Bigg]^2 -
\Bigg[\mu \, M \, + \, \frac{\sqrt{\lambda}}{2 \pi}\, \sqrt{\frac{2\, \chi}{v}-1-\chi^2} \Bigg]^2 = J^2 \, . 
\end{equation}
Euclideanizing \eqref{ansatz-NS} 
and rewriting it in Poincare coordinates we have
\begin{eqnarray} \label{ansatz-NS-Poincare}
&&
\mathbf{x} \, = \, \frac{\mathbf{X}}{2}\,\tanh \left(\kappa \tau - i \, T_y\right)\, , 
\quad 
x^+ \, = \, i \, \frac{\mathbf T}{2}\,\tanh \left(\kappa \tau - i \, T_y\right) \, , 
\nonumber \\[5pt]
&& x^- \, = \, - \, \frac{i}{2} \, \left[\frac{\kappa }{m}\,Z_y^2 \, - \, 
\frac{\mathbf{X}^2}{2\, \mathbf T}\right]\,\tanh \left(\kappa \tau - i \, T_y\right) \, + \, i \, \mu^2 \, m \, \tau + V_y \, ,
\nonumber \\[5pt]
&&
\varphi_1\,=\, i \, \omega \,\tau\, + \, \Phi_y \, ,
\quad 
z \, = \, \sqrt{\frac{\kappa \, \mathbf T}{2 \, m}}\, 
\frac{Z_y}{\cosh \left(\kappa \tau - i \, T_y\right) }\, ,
\end{eqnarray}
where now $y$ is defined in \eqref{definition-y-poincare}. 
Computing the Lagrangian on shell, we finally have that 
\begin{equation} \label{Lag-onshell-NS}
{\cal L}_{\text{on shell}}^{\text{NS}} \, = \, - \, i \, \frac{\sqrt{\lambda}}{2 \, \pi}\,
\frac{Z_y^2-1}{Z_y^2}  \, \frac{\chi \left(\chi - v\right)}{1-v^2} \, \omega^2\, .
\end{equation}



\section{Three-point correlation functions}

In this section we will evaluate the leading term in the semiclassical expansion of the three point function involving 
the dilaton and two {\it heavy} operators. The two {\it heavy} operators will be dual to each of the classical solutions 
presented in the previous section. 

In general, as discussed in  \cite{Zarembo:2010rr,Costa:2010rz,Roiban:2010fe} the ratio of the three-point correlator 
of a {\it light} state and two {\it heavy} ones over the two-point function of the {\it heavy} states is given 
in the semiclassical limit by
\begin{equation} \label{recipe0}
\frac{G_{3}(\bar{x}_1, \bar{x}_2, \bar{x}_3)}{G_{2}(\bar{x}_1, \bar{x}_2)} \, = 
\int d^2 \sigma V_L\left(z(\tau,\sigma), \, \bar{x}(\tau,\sigma) - \bar{x}_{3}, \, x^{-}(\tau,\sigma), \, X^i(\tau,\sigma)\right)\,  . 
\end{equation}
Here $V_L$ is the vertex operator of the {\it light} state and $z(\tau,\sigma)$, $\bar{x}(\tau,\sigma)$, 
$x^{-}(\tau,\sigma)$ and $X^{i}(\tau,\sigma)$ with $i=1,\ldots 5$ denote the ten coordinates of the classical string solution 
tunneling from the point $\bar{x}_1$ of the boundary -which is at $z=0$- to another point $\bar{x}_2$ of the boundary, where $\bar{x}_i=(t_i,\vec{x}_i)$. 
This tunneling solution is the string theory dual of  the corresponding field theory two point correlator at strong coupling.

The vertex operator of a generic {\it light} state will depend on the deformation parameter $\mu$ and its form is not 
known for the Schrodinger background we are using. It would be interesting to construct systematically these vertex 
operators. However, there is one case where the vertex operator is actually known, even for the deformed background. 
This is the case of the dilaton since the latter couples universally to all other fields of the theory through the term $\int d^2 \sigma e^{\frac{\phi(X)}{2}}G_{\mu \nu}(X) \partial X^{\mu} \bar{\partial}X^{\nu}+...$    . 
In this case the ratio of the three to the two-point function takes the form 
\begin{equation} \label{recipe}
\frac{G_{3}(\bar{x}_1, \bar{x}_2, \bar{x}_3)}{G_{2}(\bar{x}_1, \bar{x}_2)} \,  =
\int d^2 \sigma \, e^{i \, M_3 \, x^{-}\left(\tau, \sigma\right)} \, K(\bar{x}_{\rm classical}(\tau,\sigma); \bar{x}_3) \,
 {\cal L}_{\text{on shell}}^{\rm classical}
\end{equation}
where all quantities appearing in the integrand are evaluated on the classical solution sourced by the two {\it heavy} 
state vertex operators. It is this classical solution that correspond to the two-point correlator. The exponential in \eqref{recipe} comes from the fact that the 
dilaton is expanded as a sum of the eigenfunctions of the mass operator a.k.a. number operator which is a good 
quantum number in the Schrodinger backgrounds. The relevant for the dilaton expansion is
\begin{equation}\label{dilaton}
\phi(z,\bar{x},x^{-}) \, = \, \sum_{M_3} e^{i \, M_3 \, x^{-}} \, \phi_{M_3}(z,\bar{x}) \, . 
\end{equation}
Furthermore, $K(\bar{x}_{\rm classical}(\tau,\sigma); \bar{x}_3)$ is the bulk to boundary propagator of the specific mode of the dilaton which has momentum $M_3$ in the $x^-$ direction and ${\cal L}_{\text{on shell}}^{\rm classical} $ is the Lagrangian density of the string which is evaluated on the classical two-point solution.

An important comment is in order. Would one like to calculate the three-point correlator not for a single mode characterized by $M_3$ but for the full dilaton field appearing in the left hand side of \eqref{dilaton} one should be able to evaluate the three-point correlator for each mode separately  and subsequently sum over $M_3$. This means that one should evaluate the three-point correlator for values of  $M_3$ that are comparable and bigger than the mass eigenstate $M$ of the {\it heavy} state. But in such a situation the approximation $M_3<<M$ does not hold any more and the method that we use in this paper to calculate the three-point function is no more applicable since the {\it light} state can no longer be considered as a small perturbation and its absorption from the {\it heavy} state will considerably alter the latter.  As a result, this fact forbids one to check the expression relating the three-point function coefficient to the derivative of the scaling dimension w.r.t. the coupling $\lambda$ which holds in the original AdS/CFT scenario unless one is able to compute the correlators for arbitrary values of $M_3$. We believe that the aforementioned relation should hold in the present case also and it would be interesting to perform the  calculation of the three-point function for arbitrary values of $M_3$ in order to verify this.


\subsection{Schrodinger symmetry and the propagator for the scalar field}

In this subsection, we briefly review the analysis presented in \cite{Fuertes:2009ex,Volovich:2009yh}. 
In a quantum field theory possessing Schrodinger symmetry the two-point function of scalar operators is completely 
determined up to an overall constant and takes the following form
\begin{equation} \label{2-point}
G_{2}(\bar{x}_1, \bar{x}_2)
= C\, \delta_{\Delta_1, \Delta_2}^{M_1, M_2} \, 
\frac{\theta(t_1-t_2)}{(t_1-t_2)^{\Delta_1}} \, \exp \left[\frac{i\, M_1}{2}\frac{(\vec{x}_1-\vec{x}_2)^{2}}{t_1-t_2}
\right]
\end{equation}
where $\bar{x}_i=(t_i,\vec{x}_i)$ and ($M_i$, $\Delta_{i}$) denote the non-relativistic mass and the scaling
dimension of the scalar operator. 

When we have three spacetime points, namely $\bar{x}_1$, $\bar{x}_2$ and $\bar{x}_3$, 
it is possible to construct the following kinematic Schrodinger invariant
\begin{equation} \label{v12}
v_{12}
=-{1 \over 2}
\left(\frac{x^2_{12}}{t_{12}}+\frac{x^2_{23}}{t_{23}}-\frac{x^2_{13}}{t_{13}} \right)
\end{equation}
where $t_{ij} = t_i-t_j$ and $x_{ij}=(\vec{x}_i-\vec{x}_j)$.
Then, the functional form of the three-point function of scalar operators is fixed up to a scaling function 
$F(v_{12})$ of this Schrodinger invariant
 \begin{eqnarray} \label{3-point}
G_{3}(\bar{x}_1, \bar{x}_2, \bar{x}_3) & = & 
\delta_{M_1+M_2}^{M_3}\,  \theta(t_1-t_3) \, \theta(t_2-t_3) \, 
t_{13}^{-{1 \over 2}\Delta_{13,2}}
t_{23}^{-{1 \over 2}\Delta_{23,1}}
t_{12}^{-{1 \over 2}\Delta_{12,3}} 
\times \nonumber \\[5pt]
&&\exp\left[\frac{ i \, M_1}{2}\frac{(\vec{x}_1-\vec{x}_3)^2}{t_1-t_3} 
+ \frac{ i \, M_2}{2}\frac{(\vec{x}_2-\vec{x}_3)^2}{t_2-t_3} \right] F(v_{12})
\end{eqnarray}
where we have introduced the following notation $\Delta_{ij,k}=\Delta_i + \Delta_{j}-\Delta_{k}$.  This is to be contrasted with 
the case of the original $AdS/CFT$ scenario where the spacetime dependence of a three-point correlator 
involving scalar fields is fully determined by conformal invariance.

We aim in computing the three-point function between two {\it heavy} operators and a {\it light} scalar operator. 
The {\it heavy} operators will be located at the spacetime points $\bar{x}_1$ and $\bar{x}_2$
while the {\it light} operator at the spacetime point $\bar{x}_3$.  In order to simplify the computation we consider the 
following assumption regarding the time ordering of the operators
\begin{equation} \label{t3infty}
t_3 < t_1 < t_2 \quad \& \quad t_3 \rightarrow - \, \infty \, .
\end{equation}
Furthermore for the semiclassical approximation to make sense we impose the condition that the mass eigenvalue of the {\it light} should be much less that the corresponding mass eigenvalue of the {\it heavy} states, that is
\begin{equation} \label{M3=0}
M_3 \ll M_1 \,\, \& \,\, M_2 \quad \xRightarrow[\text{}]{\delta_{M_1+M_2}^{M_3}} \quad M_1 \approx - M_2 \, . 
\end{equation}
Normalizing the three-point function of \eqref{3-point} by dividing with the two-point function \eqref{2-point} and
using \eqref{t3infty}, \eqref{M3=0} and $\Delta_1 = \Delta_2$ we arrive to following expression
\begin{eqnarray} \label{3over2-point-function}
\frac{G_{3}(\bar{x}_1, \bar{x}_2, \bar{x}_3)}{G_{2}(\bar{x}_1, \bar{x}_2)} & \approx &
\left( \frac{t_{12}}{t_{13} \, t_{23}}\right)^{\Delta_3 \over 2} \, \exp \big[ - i \, M_1 \, v_{12} \big] \, F(v_{12})
\nonumber \\
& \approx & 
\left(\frac{\sqrt{t_{21}}}{- \,  t_3}\right)^{\Delta_3} \, \tilde{F}\left(v_{12}\right),
\end{eqnarray}
where in the last equality we have used the fact that $t_3<<t_1<t_2$.
Furthermore, $\Delta_3$ is the conformal dimension of the {\it light} operator
\begin{equation}
\Delta_3 \, = \, 2 \, + \, \sqrt{4\, + \,  \mu^2\, M_3^2} \, 
\end{equation}
and
\begin{equation} \label{ftilde}
 \tilde{F}(v_{12}) \,= \,  \, \exp \big[ - i \, M_1 \, v_{12} \big] \, F(v_{12}) \, . 
\end{equation}
The strong coupling computation should verify the spacetime structure of \eqref{3over2-point-function} and 
predict the expression of the scaling function $\tilde{F}\left(v_{12}\right) $. 

The bulk to boundary propagator for the scalar field, here the dilaton modes,  is given by the following expression
\begin{eqnarray} \label{propagator}
K(z, \vec{x},t; \vec{x}_3, t_3) & = &
\frac{i( \frac{\mu \, M_3}{2})^{\Delta_3-1}e^{-{i \over 2}\pi \Delta_3}}{\pi \, \Gamma(\Delta_3-2)} \,  \theta(t-t_3)
\left(\frac{z}{t-t_3}\right)^{\Delta_3}
e^{{i\over 2} \, \mu \, M_3 {\frac{z^2+(x-x_3)^2}{(t-t_3)}}}
\nonumber \\
& \approx & \frac{i\left( \frac{\mu \, M_3}{2}\right)^{\Delta_3-1}e^{-{i \over 2}\pi \Delta_3}}{\pi \, \Gamma(\Delta_3-2)} 
\left(\frac{z}{- \, t_3}\right)^{\Delta_3}
\end{eqnarray}
where to obtain the second line of \eqref{propagator} we have used \eqref{t3infty}. It is exactly that 
expression of the propagator that we will use in the next subsections to perform the three-point function 
calculation. 

\subsection{Three-point function of dilaton with giant magnons}
\label{3point-GM}

Before presenting the actual computation, we collect the essential components for the calculation of the three-point function 
of two giant magnons, either dyonic or single spin, and the operator dual to the dilaton.
 In particular,  we rewrite in Poincare coordinates the dyonic and single spin giant magnon solutions 
of \eqref{ansatz-GM-SS-Poincare} and  \eqref{ansatz-GM-Poincare} respectively by
using ratios of conserved quantities\footnote{Notice that even if both $E$ and $M$ are infinite quantities, their 
ratio is finite.}
\begin{equation}
z = \sqrt{\frac{\mathbf T}{2} \left(\frac{E}{M} + \frac{v_{12}}{2} \right)}\, 
\frac{1}{\cosh\kappa \tau }
\quad \& \quad 
x^- = - \frac{i}{2} \, \left(\frac{E}{M} - \frac{v_{12}}{2} \right) \tanh\kappa \tau + i  \mu^2 \, m  \, \tau  +  V_y(y) 
\end{equation}
where the function $V_y(y)$\footnote{To obtain \eqref{Vy-dyonic-GM} one needs to integrate $V_y'(y)$, whose explicit 
expression appears in \cite{Georgiou:2017pvi}. We have fixed the unimportant constant of integration by setting $V_y(-\infty) = 0$. } 
is given by the following expression
\begin{eqnarray} \label{Vy-dyonic-GM}
V_y(y) \,& =& - \mu \, \sqrt{\frac{\omega_{\phi}}{\omega_{\psi} + 2 \, \mu \, m \, v} }\, \sin \frac{p}{2} \left[\tanh\big(\sqrt{\beta_4}\, y\big) +1\right]
\nonumber \\[5pt]
&=& - \mu \, \delta \, \sin \frac{p}{2} \left[\tanh\big(\sqrt{\beta_4}\, y\big) +1\right] \, . 
\end{eqnarray}
In the last line of \eqref{Vy-dyonic-GM} we have used the fact that
\begin{equation} \label{def-delta}
\sqrt{\frac{\omega_{\phi}}{\omega_{\psi} + 2 \, \mu \, m \, v} } \, = \,  
\Bigg[\frac{J_2}{2\, T\, \sin \frac{p}{2}} + \frac{1}{2} \sqrt{\frac{J_2^2}{T^2\, \sin^2 \frac{p}{2}} +4}\Bigg]^{-1}
\equiv \delta
\end{equation}
to write the quantity under the square root as a function of conserved quantities.  
Note that $J_2$ is a finite quantity. 
Setting $J_2=0$ (or equivalently $\omega_{\phi} = \omega_{\psi} + 2 \, \mu \, m \, v$) in \eqref{def-delta} fixes 
$\delta=1$ and from \eqref{Vy-dyonic-GM} we obtain the expression for the function $V_y(y)$ in the case 
of the single spin giant magnon solution. Furthermore, we have used the fact that $t_3 \rightarrow -\infty$ to write 
$\frac{\mathbf{X}^2}{2\, \mathbf T}$ as $v_{12}$ (see \eqref{v12} and keep in mind that $t_{21}=\mathbf T$ and 
$x^2_{12}=\mathbf{X}^2$). At this point it should be apparent why we had 
to generalize the solutions of \cite{Georgiou:2017pvi}. 
Those solutions have $\bold X=0$ and this will make the invariant $v_{12}=0$. Had we sticked to the solutions of \cite{Georgiou:2017pvi} all the dependence of the three-point functions on $v_{12}$ would have been lost.  

Now we are in the position to write the expression for the normalized three-point function of the dyonic 
giant magnon by inserting the appropriate solution in the general expression for the 
three-point correlator \eqref{recipe} to get
\begin{equation} \label{3point-GM-v1}
\frac{G_{3}}{G_{2}} = - \, i \, \frac{\sqrt{\lambda}}{2 \, \pi}\,
\frac{\omega_{\phi}\, \omega_{\psi} }{1 - v^2} \, \left(1+\frac{\mu\, m \, v}{ \omega_{\psi}}\right)\,
\frac{i\left( \frac{\mu \, M_3}{2}\right)^{\Delta_3-1}e^{-{i \over 2}\pi \Delta_3}}{2^{\frac{\Delta_3}{2}} 
\pi \, \Gamma(\Delta_3-2)} 
 \, \left[\frac{E}{M} + \frac{v_{12}}{2} \right]^{\frac{\Delta_3}{2}} 
\left(\frac{\sqrt{t_{21}}}{-\, t_3}\right)^{\Delta_3}\, {\cal I}_{\rm GM}
\end{equation}
where ${\cal I}_{\rm GM}$ is given by the following double integral\footnote{For the dyonic giant magnon the 
expression for the function $u(y)$ can be found in \cite{Georgiou:2017pvi}, while for the single spin giant magnon
it is given in \eqref{u-solution}.}
\begin{equation}
{\cal I}_{\rm GM} \, = \, \frac{1}{c}\, \int d\tau \, dy \frac{e^{i \, M_3 \, x^{-}}}{\left(\cosh\kappa \tau\right)^{\Delta_3}} \, u(y) \, . 
\end{equation}
The two integrals can be performed independently and we can write
\begin{equation} \label{I-GM}
{\cal I}_{\rm GM} \, = {\cal I}_1 \, {\cal I}_2 
\end{equation}
with 
\begin{equation} \label{I1-GM-v1}
{\cal I}_1 \, = \, \int^{+\infty}_{-\infty} \frac{\exp 
\Big[\frac{M_3}{2} \left(\frac{E}{M} - \frac{v_{12}}{2} \right) \,\tanh\kappa \, \tau 
- \mu^2 \, m \, M_3 \, \tau\Big]}
{\left(\cosh \kappa\, \tau \right)^{\Delta_3} } \, d\tau
\end{equation}
and
\begin{equation} \label{I2-GM}
{\cal I}_2 = \frac{1}{c} \, \int_{-\infty}^{+\infty}u(y)\, e^{i \, M_3 \, V_y(y)}\, dy \, = \, 
 \frac{i}{M_3}\, \frac{\left(1-v^2\right)}{\mu \, \omega_{\phi}} \left(e^{- \,2 \, i \, \mu \, M_3\, \sin \frac{p}{2} } -1\right) \, . 
\end{equation}
The integral in \eqref{I1-GM-v1} can be evaluated analytically using the approximation $\tanh y \approx y$ 
\begin{equation} \label{I1-GM-v2}
{\cal I}_1 \approx \frac{2^{\Delta_3 -1}}{\kappa} \,
\mathrm{B} \left[\frac{1}{2} \left(\Delta_3 - \Xi_{\rm GM}\right), \frac{1}{2} \left(\Delta_3 +\Xi_{\rm GM}\right)\right]
\end{equation}
with 
\begin{equation}
\Xi_{\rm GM} \, = \, \frac{1}{2} \, M_3 \left(\frac{E}{M} - \frac{v_{12}}{2} \right) \, - \, \mu^2 \, M_3 \, 
\left(\frac{E}{M} + \frac{v_{12}}{2} \right)^{-1}\, .
\end{equation}
Combining \eqref{3point-GM-v1} with \eqref{I-GM}, \eqref{I2-GM} and \eqref{I1-GM-v2} we arrive to the 
following expression for the three-point coupling for the dyonic giant magnon case
\begin{eqnarray}  \label{3point-GM-2s} 
{\tilde F}(v_{12}) & = & i \, 
\frac{\sqrt{\lambda}}{2\, \pi^2}\, \mu^{\Delta_3-2}\, M_3^{\Delta_3-2} \, 
\frac{e^{-{i \over 2}\pi \Delta_3}}{2^{\frac{\Delta_3}{2}} \Gamma(\Delta_3-2)} \, 
\left(\frac{E}{M} + \frac{v_{12}}{2}\right)^{\frac{\Delta_3}{2}} 
\nonumber  \\[5pt] 
&&
\sqrt{1-\frac{\mu^2}{\left(\frac{E}{M} + \frac{v_{12}}{2}\right)^{2}}} \, 
\Bigg[ 1 \, + \, \frac{\left(1 \, - \, \delta \right)\, \left(1 \, - \, v^2\right)}{1\, +\, \delta \,+\,  2 \, v \, \sqrt{\delta} \, 
\cos \frac{p}{2}}\Bigg]
\nonumber  \\ [5pt]
&& \mathrm{B} \left[\frac{1}{2} \left(\Delta_3 - \Xi_{\rm GM} \right), \frac{1}{2} \left(\Delta_3 + \Xi_{\rm GM} \right)\right]
\, \left(e^{- \,2 \, i \, \mu \, M_3\, \delta \, \sin \frac{p}{2} } -1\right)
\end{eqnarray}
where ${\tilde F}$ is the scaling function defined in \eqref{ftilde}. The Schrodinger invariant ratio $v_{12}$ appears in several places in \eqref{3point-GM-2s}. It appears in the beta function through its dependence on $\Xi$, as well as 
in the part of the prefactor $(\frac{E}{M} + \frac{v_{12}}{2})^{\frac{\Delta_3}{2}}$. 
This is the string prediction for the leading term in the large $\lambda$ expansion of the scaling function ${\tilde F}$.
Notice that the spacetime dependence in \eqref{3point-GM-v1} is the one 
dictated by non-relativistic conformal invariance \eqref{3over2-point-function}. 
%
In order to express the ratio $\frac{\mu \, m}{\omega_{\phi}}$ as a function of conserved quantities
we have used the following expression\footnote{In order to derive \eqref{sinp-dyonic} we write 
$\sin^2 \frac{p}{2} = \frac{\beta_4}{|\beta_6|}$ and substitute the values of the constants from \cite{Georgiou:2017pvi}.}
\begin{equation} \label{sinp-dyonic}
\sin^2 \frac{p}{2} = -1 + \frac{1}{4 \, \delta} \, \left[ v \left(1+\delta \right) + 2 \, \frac{\delta \,\mu \, m}{\omega_{\phi}} \, 
\left(1-v^2\right)\right]^2 \, . 
\end{equation}
Finally, notice that in the dyonic giant magnon solution there are two finite conserved quantities, namely $p$ and $J_2$, 
and we have expressed the three-point coupling in terms of those quantities, together with the finite ratio of $E$ and $M$. 

Setting $\delta =1$ (or equivalently $J_2=0$) we obtain the three-point coupling for the single spin giant magnon case
\begin{eqnarray}  \label{3point-GM-1spin} 
{\tilde F}(v_{12}) & = & i \, 
\frac{\sqrt{\lambda}}{2\, \pi^2}\, \mu^{\Delta_3-2}\, M_3^{\Delta_3-2} \, 
\frac{e^{-{i \over 2}\pi \Delta_3}}{2^{\frac{\Delta_3}{2}} \Gamma(\Delta_3-2)} \, 
\left(\frac{E}{M} + \frac{v_{12}}{2}\right)^{\frac{\Delta_3}{2}} \, 
\sqrt{1-\frac{\mu^2}{\left(\frac{E}{M} + \frac{v_{12}}{2}\right)^{2}}} 
\nonumber  \\[5pt] 
&& \mathrm{B} \left[\frac{1}{2} \left(\Delta_3 - \Xi_{\rm GM} \right), \frac{1}{2} \left(\Delta_3 + \Xi_{\rm GM} \right)\right]
\, \left(e^{- \,2 \, i \, \mu \, M_3 \, \sin \frac{p}{2} } -1\right) \, . 
\end{eqnarray}


\subsection{Three-point function of dilaton with single spikes}
\label{3point-SS}

The calculation of the three-point function for the single spike solution (either dyonic 
\eqref{ansatz-GM-SS-Poincare} or single spin \eqref{ansatz-SS-Poincare}) has to be performed with 
special care. Integration of the on-shell Lagrangian (either in \eqref{Lag-onshell-SS-2spin} or in \eqref{Lag-onshell-SS}) 
will lead to an infinity and for this reason 
a proper normalization has to be implemented. The most natural/physical one is to subtract in both cases 
the Lagrangian for a classical configuration that moves very close to the equator, i.e. 
$\theta = \pi $ $\Rightarrow$ $u=0$ which represents a hoop winding an infinite number of times the equator of $S^5$. 
The remaining on-shell Lagrangians for the 
three-point function calculation become
\begin{equation} \label{Lag-onshell-SS-2spin-v2}
{\cal L}_{\text{on shell}}^{\text{2s-norm}} \, = \, i \, \frac{\sqrt{\lambda}}{2 \, \pi}
 \, \frac{v \, \omega_{\phi}}{2 \, \mu} \,
\Big[2 \, \alpha - v \, \mu \, \left(\omega_{\phi} -\omega_{\psi} \right)\Big] \, \frac{u(y)}{1-v^2}
\end{equation}
and
\begin{equation} \label{Lag-onshell-SS-v2}
{\cal L}_{\text{on shell}}^{\text{1s-norm}} \, =  \, i \, \frac{\sqrt{\lambda}}{2 \, \pi}\,
\frac{1}{2} \, \frac{\alpha^2}{\mu^2}  \, \left(1+ \frac{\mu \, v \, \omega}{\alpha} \right) \, \frac{u(y)}{1-v^2} \, . 
\end{equation}

After properly normalizing the Lagrangians, the computation of the three-point function, where instead of giant magnons 
we have single spikes, is similar to the calculation we presented in the previous subsection. Now the function $V_y(y)$
is given by
\begin{equation} \label{Vy-dyonic-SS}
V_y(y) \, = \,  \frac{\mu \,\pi \left(J_1 +J_2\right)}{\,\sqrt{\lambda}} \,\left[\tanh\big(\sqrt{\beta_4}\, y\big) +1\right] \, . 
\end{equation}
Setting $J_2=0$ (or equivalently $\omega_{\phi}= \frac{\alpha}{\mu \, v }$) we obtain the expression for 
the function $V_y(y)$ in the case of the single spin single spike solution.
Substituting all the ingredients in \eqref{recipe} we arrive to a double integral (in $\tau$ and $y$)
for the three-point function, similar to that of \eqref{I-GM}. In the dyonic spike case the calculation of those
integrals give the following results
\begin{equation}  \label{I1-SS}
{\cal I}_1 \approx \frac{2^{\Delta_3 -1}}{\kappa} \,
\mathrm{B} \left[\frac{1}{2} \left(\Delta_3 - \Xi_{\rm SS}\right), \frac{1}{2} \left(\Delta_3 +\Xi_{\rm SS}\right)\right]
\end{equation}
with
\begin{equation}
\Xi_{\rm SS} \, = \, \frac{1}{2} \, M_3 \left(\frac{E}{M} \, - \, \frac{v_{12}}{2}\right)  - \mu \, M_3
\end{equation}
and
\begin{equation}  \label{I2-SS}
{\cal I}_2 = \frac{i}{M_3}\, \frac{c \, \left(1-v^2\right)}{\mu \, v^2 \, \omega_{\phi}} \left(1 - e^{i \, \mu \, M_3\, 
\frac{2 \, \pi(J_1 +J_2)}{\sqrt{\lambda}} }\right) \, . 
\end{equation}
The three-point coupling for the dyonic single spike case is given by the following expression
\begin{eqnarray} \label{3point-SS-2s} 
{\tilde F}(v_{12}) & = &-\,  i \, 
\frac{\sqrt{\lambda}}{2\, \pi^2 \, v}\, \mu^{\Delta_3-2}\, M_3^{\Delta_3-2} \, 
\frac{e^{-{i \over 2}\pi \Delta_3}}{2^{\frac{\Delta_3}{2}} \, \Gamma(\Delta_3-2)}
\left(\frac{E}{M} + \frac{v_{12}}{2}\right)^{\frac{\Delta_3}{2}}
\left(1-\frac{\mu}{\frac{E}{M} + \frac{v_{12}}{2}} - \frac{J_2}{J_1}\right)
\nonumber  \\ [5pt]
&& \mathrm{B} \left[\frac{1}{2} \left(\Delta_3 - \Xi_{\rm SS} \right), \frac{1}{2} \left(\Delta_3 + \Xi_{\rm SS} \right)\right]
\, \left(1 - e^{i \, \mu \, M_3\, \frac{2 \, \pi (J_1 +J_2)}{\sqrt{\lambda}} }\right) \, . 
\end{eqnarray}
In order to write the three-point coupling in \eqref{3point-SS-2s} using conserved quantities, we have used the 
following expressions
\begin{equation}
1 - \frac{\mu}{\frac{E}{M} + \frac{v_{12}}{2}} = \frac{\mu \, v \left(\omega_{\psi} + \omega_{\phi}\right)}{2 \, \alpha}
\quad \& \quad 
\frac{J_2}{J_1} = \frac{\mu \, v \, \omega_{\phi} - \alpha}{\alpha} \, . 
\end{equation}
Notice that in the dyonic single spike solution there are two finite conserved quantities, namely $J_1$ and $J_2$, 
and we have expressed the three-point coupling in terms of those quantities, together with the finite ratio of E and M.

Setting $J_2=0$ in \eqref{3point-SS-2s} we obtain the three-point coupling for the single spin single spike case 
\begin{eqnarray} \label{3point-SS-1s} 
{\tilde F}(v_{12}) & = &-\,  i \, 
\frac{\sqrt{\lambda}}{2\, \pi^2 \, v}\, \mu^{\Delta_3-2}\, M_3^{\Delta_3-2} \, 
\frac{e^{-{i \over 2}\pi \Delta_3}}{2^{\frac{\Delta_3}{2}}  \Gamma(\Delta_3-2)}
\left(\frac{E}{M} + \frac{v_{12}}{2}\right)^{\frac{\Delta_3}{2}}
\left(1-\frac{\mu}{\frac{E}{M} + \frac{v_{12}}{2}}\right)
\nonumber  \\ 
&& \mathrm{B} \left[\frac{1}{2} \left(\Delta_3 - \Xi_{\rm SS} \right), \frac{1}{2} \left(\Delta_3 + \Xi_{\rm SS} \right)\right]
\, \left(1 - e^{i \, \mu \, M_3\, \frac{ 2 \, \pi\,J}{\sqrt{\lambda}} }\right) \, . 
\end{eqnarray}


\subsection{Three-point function of dilaton with spinning BMN-like strings}
\label{3point-BMN}

In the case of the spinning BMN-like strings the computation of the three-point function can be easily performed 
using the strategy we described in subsection \ref{3point-GM}. For that reason we only quote the results for the 
calculation of the integrals ${\cal I}_1$
\begin{equation}
{\cal I}_1 \approx \frac{2^{\Delta_3 -1} \, \sqrt{\lambda}}{E} \,
\mathrm{B} \left[\frac{1}{2} \left(\Delta_3 - \Xi_{\rm BMN}\right), \frac{1}{2} \left(\Delta_3 + \Xi_{\rm BMN}\right)\right]
\end{equation}
with
\begin{equation}
\Xi_{\rm BMN} \, = \, \frac{1}{2} \, M_3 \left(\frac{E}{M} \, - \, \frac{u_{12}}{2}\right)  - \mu \, M_3 \, 
\sqrt{1\, - \, \frac{J^2}{\left(E \, + \, \frac{u_{12}}{2} \, M\right)^2}}
\end{equation}
and ${\cal I}_2$
\begin{equation} \label{integral-tau}
{\cal I}_2 \, = \, \int^{2 \,\pi}_{0} e^{- i \, M_3 \, \xi \, \frac{J}{M} \, \sigma}\, d\sigma \, = \, 
\, \frac{i}{M_3 \, \xi} \frac{M}{J}\, \left( e^{- 2 \, \pi \, i \, M_3 \, \xi \, \frac{J}{M} } -1\right) \, . 
\end{equation}
Notice that in the spinning BMN-like string solution all the conserved quantities are finite.
The three-point coupling is given by the following expression 
\begin{eqnarray} \label{3point-BMN}
{\tilde F}(v_{12}) & = & i \, 
\frac{\sqrt{\lambda}}{2\, \pi^2}\, \mu^{\Delta_3-1}\, M_3^{\Delta_3-2} \, 
\frac{e^{-{i \over 2}\pi \Delta_3}}{2^{\frac{\Delta_3}{2}}\,  \Gamma(\Delta_3-2)} \, 
\left(\frac{E}{M} + \frac{u_{12}}{2} \right)^{\frac{\Delta_3}{2}} \, 
\frac{M}{J} \, \sqrt{1-\frac{J^2}{\left(E \, + \, \frac{u_{12}}{2} \, M\right)^2}}
\nonumber  \\ 
&& \mathrm{B} \left[\frac{1}{2} \left(\Delta_3 - \Xi_{\rm BMN} \right), \frac{1}{2} \left(\Delta_3 +\Xi_{\rm BMN}\right)\right]
\, \left( e^{- 2 \, \pi \, i \, M_3 \, \xi \, \frac{J}{M} } -1\right) \, . 
\end{eqnarray}


\subsection{Three-point function of dilaton with new classical string solution}
\label{3point-NS}

Following the same strategy an in subsection \ref{3point-GM} to present the three-point coupling calculation, 
we rewrite the essential components of the Poincare coordinates 
in \eqref{ansatz-NS-Poincare} using ratios of conserved quantities
\begin{eqnarray}
&& x^- \, = \, - \, \frac{i}{2} \, \left[\mu\, \sqrt{1+\chi^{-2}}\,\,Z_y^2 \, - \, u_{12}\right]\,\tanh 
\left(\kappa \tau - i \, T_y\right) \, + \, i \, \mu^2 \, m \, \tau + V_y 
\nonumber \\[5pt]
&&
z \, = \, \sqrt{\frac{\mathbf T\, \mu}{2} \sqrt{1+\chi^{-2}}}\, 
\frac{Z_y}{\cosh \left(\kappa \tau - i \, T_y\right) }
\end{eqnarray}
where the expressions of $T_y$ and $V_y$ as functions of $Z_y$ can be obtained by integrating 
\eqref{constraintT-NS} and \eqref{constraintV-NS} using also \eqref{constraintZ}. Here we quote
those results, initially for the function $T_y$
\begin{equation}
\tan T_y(Z_y)  = \frac{Z_y}{\sqrt{\frac{2 \, \chi}{v\, \left(1+\chi^2\right)} - Z_y^2}}
\end{equation}
and subsequently for the function $V_y$
\begin{equation}
V_y(Z_y)  = - \frac{\mu \, Z_y }{2} \,\sqrt{1+\chi^{-2}} \sqrt{\frac{2 \, \chi}{v\, \left(1+\chi^2\right)} - Z_y^2} \, . 
\end{equation}
The connection between the quantities $\vec X_0$ and $v_{12}$ is $v_{12} = - \vec X_0^2$. 
Notice that in the new classical string solution all the conserved quantities, namely $E$ in \eqref{Energy-NS}, $M$ in
\eqref{M-NS} and $J$ in \eqref{J-NS} are infinite and in the expression of the three-point coupling only ratios of them 
may appear. 

\begin{figure}[h] 
   \centering
   \includegraphics[width=7.5cm]{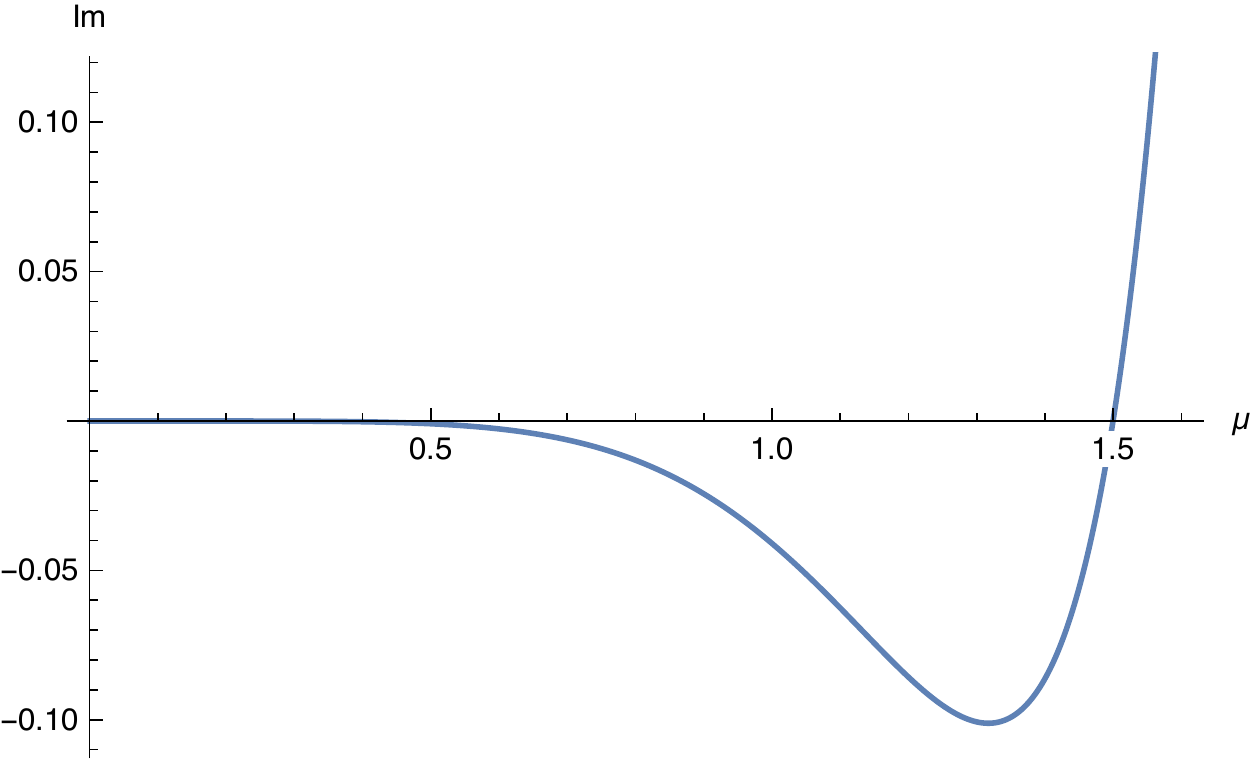}
    \includegraphics[width=7.5cm]{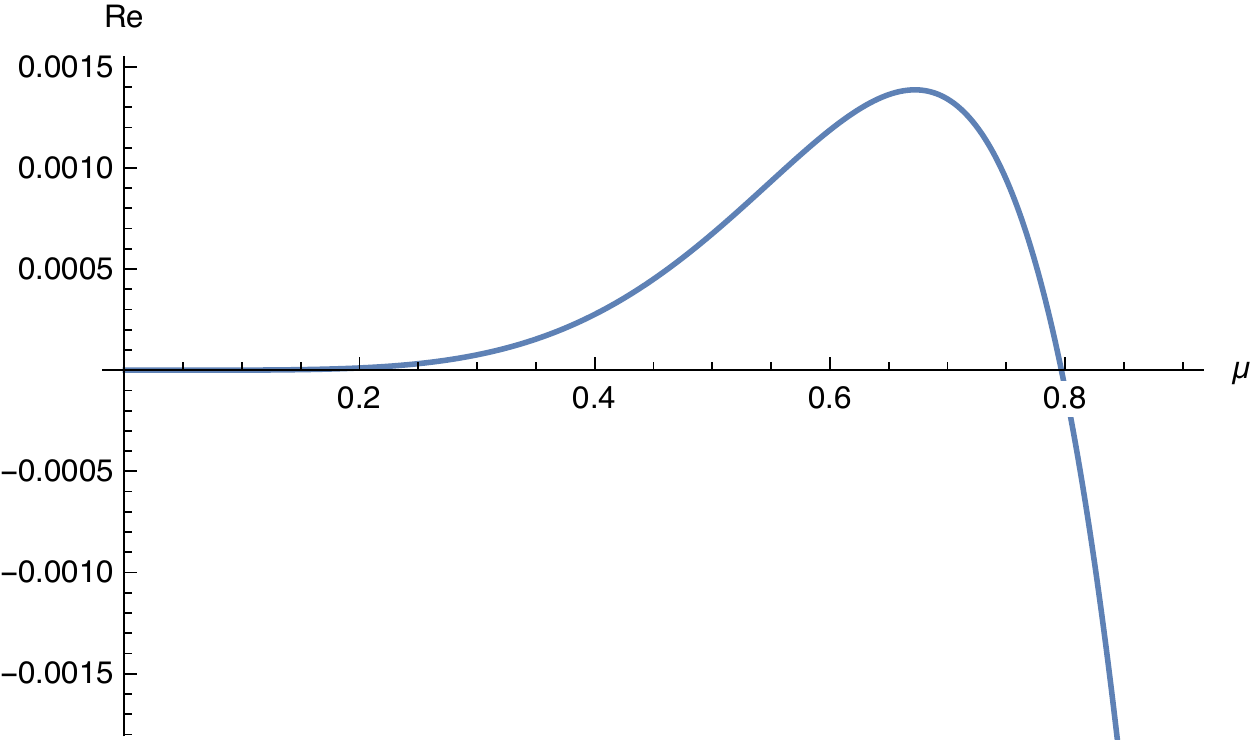}
    \includegraphics[width=7.5cm]{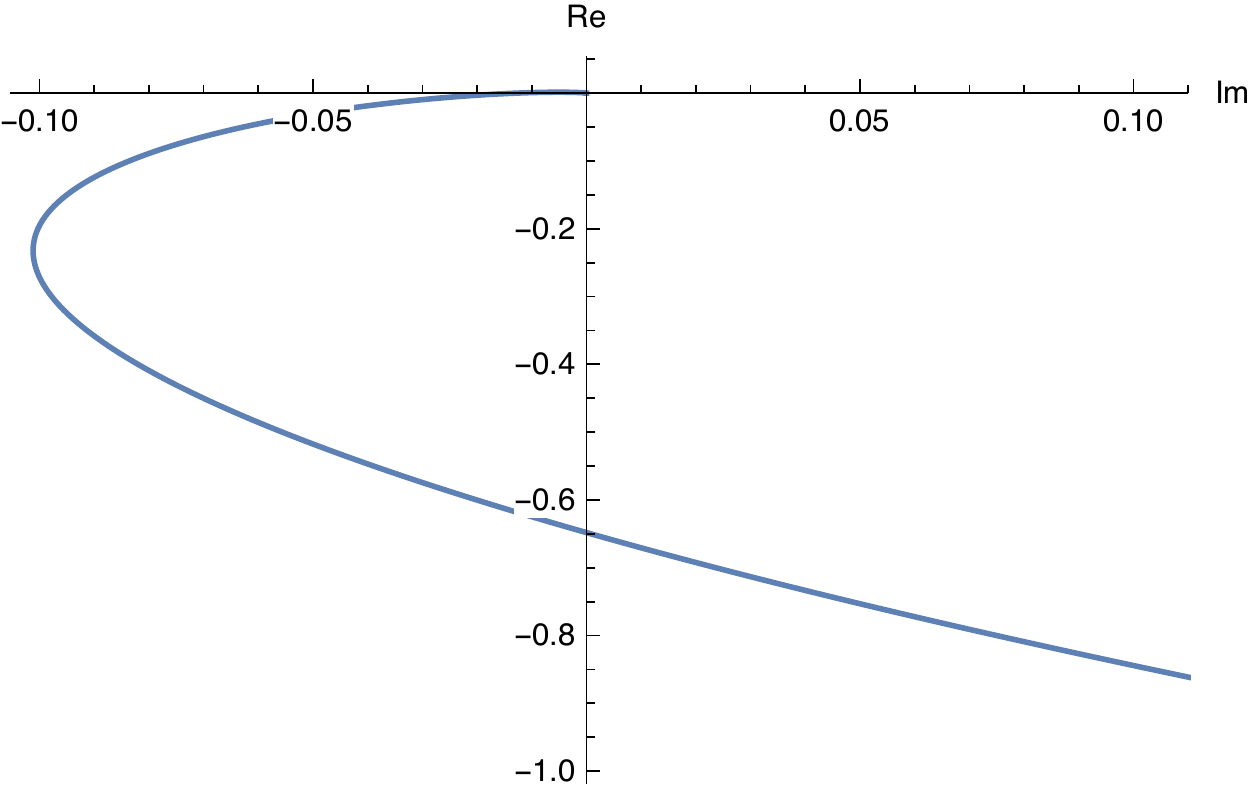}
       \caption{Plots for the three-point coupling in \eqref{3point-NS} of the new solution.
       In the upper two panels we plot the imaginary and the 
       real part of the the three-point coupling as a function of $\mu$ while in the third panel we plot the real and the 
       imaginary part of the three-point coupling for different values of the deformation parameter.}
   \label{3pointNSplots}
\end{figure}

Plugging all the necessary ingredients in \eqref{recipe}, we obtain the normalized three-point function as follows
\begin{eqnarray} 
\frac{G_{3}}{G_{2}} & = & 
\frac{\sqrt{\lambda}}{\pi} \, \frac{\mu \, m \,\left(\chi - v\right)}{v\, \sqrt{1+\chi^2}}
\frac{\left( \frac{\mu \, M_3}{2}\right)^{\Delta_3-1}e^{-{i \over 2}\pi \Delta_3}}{\pi 
\, \Gamma(\Delta_3-2)} \, \left[\frac{\mu}{2} \,\sqrt{1+\chi^{-2}}\right]^{\frac{\Delta_3}{2}}\, 
\left[\frac{\sqrt{t_{21}}}{- \, t_3}\right]^{\Delta_3} 
\nonumber \\[5pt]
&&\int^{\infty}_{-\infty} d\tau \int^{Z_{qrit}}_1 \frac{d  \, Z_y}{Z_y^{2-\Delta_3} \, \sqrt{Z_{\rm qrit}^2 \, - \, Z_y^2}} \, 
\frac{e^{i \, M_3 \, x^{-}}}{\left[\cosh \left(\kappa \tau - i \, T_y\right)\right]^{\Delta_3} }  
\end{eqnarray}
where we have changed variables, from $\sigma$ to $Z_y$ according to \eqref{change-variables}. 
The integral with respect to $\tau$ in the expression above can be calculated analytically using the 
approximation $\tanh y \approx y$ and the result is 
\begin{eqnarray} 
&&\int^{\infty}_{-\infty} d\tau \, 
\frac{e^{\frac{1}{2} \, M_3\, \left[\mu\, \sqrt{1+\chi^{-2}}\,\,Z_y^2 \, - \, u_{12}\right]\,\tanh \left(\kappa \tau - i \, T_y\right) \, -  \, \mu^2 \, M_3 \, m \, \tau}}{\left[\cosh \left(\kappa \tau - i \, T_y\right)\right]^{\Delta_3} } \approx
\nonumber \\[5pt]
&& \frac{2^{\Delta_3 -1}}{\kappa} \, e^{- i \, \mu \, M_3 \, T_y \, \frac{\chi}{\sqrt{1+\chi^2}}} \, 
\mathrm{B} \left[\frac{1}{2} \left(\Delta_3 - \Xi_{\rm NS}\right), \frac{1}{2} \left(\Delta_3 +\Xi_{\rm NS}\right)\right]
\end{eqnarray}
with
\begin{equation}
\Xi_{\rm NS} = \frac{1}{2} \, M_3 \left[\mu\, \sqrt{1+\chi^{-2}}\,\,Z_y^2 \, - \, v_{12}\right] - \mu \, M_3 \, 
\frac{\chi}{\sqrt{1+\chi^2}} \, . 
\end{equation}
Contrary to the cases of the previous subsections, the integral with respect to $Z_y$ cannot be performed analytically,
since the expression for $\Xi_{NS}$ depends also on the integration variable $Z_y$.
The three-point coupling for the new classical string solution is written in terms of that integral as follows
\begin{eqnarray} \label{3point-NS}
{\tilde F}(v_{12})  & = & \frac{\sqrt{\lambda}}{\pi^2} \, \left(\mu \, M_3\right)^{\Delta_3-1} \, 
\frac{\chi \, \left(\chi - v\right)}{v\,\left(1+\chi^2\right)}
\frac{e^{-{i \over 2}\pi \Delta_3}}{\Gamma(\Delta_3-2)} \, \left[\frac{\mu}{2} \,\sqrt{1+\chi^{-2}}\right]^{\frac{\Delta_3}{2}}
\\[5pt]
&& \int^{Z_{qrit}}_1d Z_y \, \frac{e^{- i \, \mu \, M_3 \, T_y \, \frac{\chi}{\sqrt{1+\chi^2}} \, + \, i \, M_3 \, V_y}}{\xi^{2-\Delta_3} \, \sqrt{Z_{\rm qrit}^2 \, - \, Z_y^2}}  \,
\mathrm{B} \left[\frac{1}{2} \left(\Delta_3 - \Xi_{\rm NS}\right), \frac{1}{2} \left(\Delta_3 +\Xi_{\rm NS}\right)\right]  \, . 
\nonumber
\end{eqnarray}
Even if the integral with respect to $Z_y$ cannot be calculated analytically, it can be performed numerically and 
fixing the values for $v$, $\chi$, $\mu$ and $v_{12}$ as follows
\begin{equation} 
v\, =\,\frac{1}{4}\, , \quad \chi\, = \, 1\, , \quad \mu\,=\, 1 \quad \& \quad v_{12} \, = \, 0
\end{equation}
we obtain the following values for the three-point coefficient
\begin{eqnarray}
&& {\cal C}_{3}^{\rm dilaton} \, \approx \, - \, 0.02 \, - \, 0.04\, i \quad {\rm for} \quad M_3 \, = \, 1
\nonumber \\[5pt]
&& {\cal C}_{3}^{\rm dilaton} \, \approx \, - \, 0.86 \, + \, 1.63\, i \quad {\rm for} \quad M_3 \, = \, 2
\end{eqnarray}
when $M_3$ takes the values 1 and 2. 
Alternatively we could fix  $v$, $\chi$, $M_3$ and $v_{12}$ as follows 
\begin{equation} 
v\, =\,\frac{1}{4}\, , \quad \chi\, = \, 1\, , \quad M_3\,=\, 1 \quad \& \quad v_{12} \, = \, 0
\end{equation}
and obtain the three-point coupling as a function of the deformation parameter $\mu$. 
Since we can vary the deformation parameter in a continuous way it is possible to obtain plots 
for the real and the imaginary value of the three-point coupling as a function of $\mu$. These plots can are presented in figure \ref{3pointNSplots}.

\appendix 

\section{Polyakov action and a consistent truncation}
\label{Polyakov-truncation}

The Polyakov form of the string action is given by the following standard expression\footnote{The string tension 
$T_{s}$ is related to the 't Hooft coupling as $T_s=\frac{\sqrt{\lambda}}{2\pi}$.}
\begin{equation}  \label{Polyakov}
S_P \, = \, - \, {T_{s} \over 2} \int d\tau d\sigma \left( \sqrt{-h} \, h^{\alpha\beta} \, g_{\alpha\beta} \,  
- \,  \epsilon^{\alpha\beta} \, b_{\alpha\beta} \right)
\end{equation}
where 
\begin{equation} \label{g&b-definition}
g_{\alpha\beta} \, = \,  G_{MN} \, \partial_\alpha x^M \, \partial_\beta x^N  
\quad \& \quad   
b_{\alpha\beta} \, = \,  B_{MN} \partial_\alpha x^M \, \partial_\beta x^N
\end{equation} 
are the pullbacks of the metric and the $B$-field on the string worldsheet and the tensor density
$\epsilon^{\alpha\beta}$ is defined according to the convention $\epsilon^{01}=1$. 
With $h_{\alpha\beta}$ we denote the worldsheet metric and we choose the conformal gauge where
$h_{\alpha\beta}=\eta_{\alpha\beta}$.  The construction is also supplemented by the Virasoro constraints, 
which are obtained by differentiating \eqref{Polyakov} with respect to $h_{\alpha \beta}$, 
\begin{equation} \label{Virasoro}
G_{MN} \left( \partial_{\tau} x^M \, \partial_{\tau} x^N \, + \,  
 \partial_{\sigma} x^M \, \partial_{\sigma} x^N \right) \, = \,  0
\quad \& \quad 
 G_{MN}  \,\partial_{\tau} x^M \, \partial_{\sigma} x^N\, = \,  0 \, .
\end{equation}
The momentum $p_M$ that is canonically conjugate to the coordinate $x^M$ is given by 
the following expression
\begin{equation} \label{momenta}
p_M\,=\,\frac{\partial \mathcal{L}}{\partial \dot{x}^M}
\end{equation} 
where $ \dot{x}^M \equiv \partial_\tau x^M$.

The 10d $Sch_5\times S^5$ metric is given by the following expression \cite{Guica:2017mtd,Kameyama:2015ufa}
\begin{eqnarray} \label{globalsch5s5}
\frac{ds^2}{R^2}&=&- \left(1+ \frac{\mu^2}{Z^4} \right) dT^2  +  \frac{1}{Z^2} \, 
\left(2 dT dV + dZ^2 - \vec{X}^2 dT^2 + d\vec{X}^2 \right) \, 
+ \, ds_{S^5}^2
\nonumber \\
\alpha' B_2& = & R^2 \, \frac{\mu}{Z^2}\,dT \wedge \left(d\chi + \omega \right) \, , 
\qquad F_5 \, = \, 4 \, R^{-1} (\omega_{Sch_5} \, +\, \omega_{S^5}  ) \, ,
\end{eqnarray}
where $\omega_{Sch_5}$ and $\omega_{S^5} $ are the volume forms of $Sch_5$ and $S^5$ respectively.
The metric in the five-sphere is written as an $S^1$-fibration over $\mathbb{C}\text{P}^2$
\begin{equation} \label{S5-CP2}
ds^2_{\rm S^5}=\left(d\chi+\omega\right)^2 \, + \, ds^2_{\rm \mathbb{C}P^2}
\quad {\rm with} \quad
ds^2_{\rm \mathbb{C}P^2}= d\nu^2+\sin^2\nu\, 
\bigl(\Sigma_1^2+\Sigma_2^2+\cos^2\nu\,\Sigma_3^2\bigr) \, .
\end{equation}
where the $\Sigma_i ~(i=1,2,3)$ and $\omega$ are defined as follows
\begin{eqnarray}
\Sigma_1&\equiv& \frac{1}{2}(\cos\psi\, d\theta - \sin\psi\sin\theta\, d\phi)
\qquad
\Sigma_2\equiv \frac{1}{2}(\sin\psi\, d\theta + \cos\psi\sin\theta\, d\phi)
\nonumber \\[5pt]
\Sigma_3&\equiv& \frac{1}{2}(d\psi - \cos\theta\, d\phi)
\qquad {\rm and} \qquad
\omega \equiv \sin^2\nu\, \Sigma_3\,. 
\end{eqnarray}
It can be explicitly checked that the following ansatz 
\begin{eqnarray} \label{ansatz-full}
&& 
T\,=\, \kappa \,\tau \, + \, T_y(y) \, , 
\quad
V \,=\,\alpha \, \tau \,  - \, \frac{\vec X^2_0}{4} \, \sin 2 \kappa  \tau \, + \, V_y(y)\, , 
\quad 
{\vec X} \,=\, {\vec X}_0 \, \sin \kappa  \tau \, ,
\quad
Z \,=\, Z_y(y) \, ,
\nonumber \\ [5pt]
&& 
\chi \, = \, 0 \, ,
\quad 
\theta\,=\,\theta_y(y) \, , 
\quad
\psi\,=\,\omega_\psi\,\tau\,+\,\Psi_y(y) \, , 
\quad
\phi\,=\,\omega_\phi\,\tau\,+\,\Phi_y(y)
\quad \& \quad 
\nu \,= \, \frac{\pi}{2}
\end{eqnarray} 
satisfies the equations of motion coming from \eqref{Polyakov} and the 
Virasoro constraints from \eqref{Virasoro}. Equipped by the 
preceding analysis we set $\chi  = 0$ and $\nu = \frac{\pi}{2}$ in equations 
\eqref{globalsch5s5} and \eqref{S5-CP2} and obtain the 
consistent truncation ansatz of \eqref{metric} and \eqref{Bfield}.


\end{document}